\documentclass[12pt,a4paper]{article}
\usepackage{jheppub_modified}

\allowdisplaybreaks[1]
\usepackage{amsmath,bm}
\usepackage{subfigure}

\newcommand{\Slash}[1]{\ooalign{\hfil /\hfil\crcr$#1$}}
\newcommand{\pa}{\partial}

\newcommand{\ket}{\rangle }
\newcommand{\bra}{\langle }

\title{Holographic Floquet states II: \\
Floquet condensation of vector mesons in nonequilibrium phase diagram
}

\author[1]{Shunichiro Kinoshita,}
\author[2]{Keiju Murata}
\author[3]{and Takashi Oka}
\affiliation[1]{Department of Physics, Chuo University, Tokyo 112-8551, Japan}
\affiliation[2]{Keio University, 4-1-1 Hiyoshi, Yokohama 223-8521, Japan}
\affiliation[3]{
Max-Planck-Institut f\"{u}r Physik komplexer Systeme (MPI-PKS), N\"{o}thnitzer Stra\ss e 38, Dresden 01187, Germany}
\affiliation[3]{Max-Planck-Institut f\"{u}r Chemische Physik fester Stoffe (MPI-CPfS), N\"{o}thnitzer Stra\ss e 40, Dresden 01187, Germany}
\emailAdd{kinoshita@phys.chuo-u.ac.jp}
\emailAdd{keiju@phys-h.keio.ac.jp}
\emailAdd{oka@pks.mpg.de}

\abstract{%
With the aim to reveal universal features of hadronic matter and correlated Dirac insulators in strong AC-electric fields, 
we study the $\mathcal{N}=2$ supersymmetric QCD with a finite quark mass
driven by a rotating electric field $
\mathcal{E}_x+i\mathcal{E}_y= E e^{i\Omega t}$. 
The analysis is done in the holographically dual D3/D7 system in the
co-rotating frame, effectively. 
The nonequilibrium phase diagram is determined from the threshold 
electric field at which the insulator phase  breaks down to a conductive phase
due to the AC version of the Schwinger mechanism. 
The external field induces a rotating current 
$\mathcal{J}_x + i \mathcal{J}_y = J e^{i\Omega t}$
originating from vacuum polarization and dissipative current in the insulating and conductive phases respectively. 
Intriguing features are observed as the frequency $\Omega$ approaches resonance with 
the meson excitation energy $\Omega_{\rm meson}$. 
There, the threshold minimizes and a condensate of vector mesons with oscillating current exists even in the zero driving field limit. 
This state, which we call {\em Floquet condensate of vector mesons}, is 
expected to be dynamically stable realizing a non-thermal fixed point that breaks time translational and reversal symmetries.
Our finding has many similarities with exciton BEC discussed in solid state systems,
where the semiconductor is to be replaced by materials hosting gapped Dirac electrons, {\em e.g.} 3D topological insulators or bismuth. 
Vector meson Floquet condensate may also have implications in the pre-thermalized dynamics in heavy ion collision experiments.
}

\begin{document}
\maketitle

\section{Introduction and overview}

Nonequilibrium dynamics after abrupt phase transitions in strongly coupled gauge theories is of fundamental interest. 
Several processes sequentially take place after the transition. 
Imagine that we initiate the transition by destroying the ground-state vacuum using a pulsed electric field. 
Being driven by the external field, the system evolves  quantum coherently for some time. 
A prominent process here is the Schwinger effect of
quark-antiquark pair production \cite{Schwinger:1951,Heisenberg1936,Weisskopf} which can trigger a vacuum breakdown. 
If the pulse is long enough, a nonequilibrium steady state may be realized as a result of a
balance between the excitation and relaxation processes, {\em e.g.} pair annihilation. 
The interest in this paper is to study the phase transition that is realized within the
nonequilibrium steady state in the presence of a driving field. 

Our analysis is done by combining the framework of AdS/CFT correspondence~\cite{Maldacena:1997re,Gubser:1998bc,Witten:1998qj}
with the Floquet theory for periodically driven systems, subsequently to the previous work~\cite{Hashimoto:2016ize}. 
The Floquet theory is a framework to study 
steady state solutions that are {\em time periodic}. In this paper, we study a specific case where the state is realized in a rotating electric field
\begin{equation}
\mathcal{E}_x+i\mathcal{E}_y= E e^{i\Omega t}\;,
  \label{rotEbc0}
\end{equation}
with a field in the $xy$-plane. 
This simplifies the 
analysis considerably since we can obtain a homogeneous and stationary solution 
by moving to the co-rotating frame (Sec.~\ref{subsec:reduction}).
Experimentally, a strong circularly polarized laser can
approximately realize a similar electric field.%
\footnote{
Note that this ``rotating electric field'' is distinct from a
``circularly polarized electromagnetic wave'', 
a vacuum solution of the Maxwell equation, since its field does
not have a spatial modulation along $z$-direction such as $e^{-ikz}$.
}

Among the strongly coupled gauge theories that can be treated holographically, we study a minimal model 
that shows phase transitions.
The model is the $\mathcal{N}=2$ supersymmetric QCD (SQCD) 
consisting of a single flavor quark and gluons that mediates interaction
with gauge group SU$(N_c)$
at large $N_c$ and at large 't Hooft coupling $\lambda \equiv N_c g_\textrm{YM}^2$. 
The holographic dual is the D3/D7 system in the probe limit: We study the classical motion of 
a single probe D7-brane in the AdS$_5\times S^5$ spacetime~\cite{Karch}. 
In the  single flavor case $N_f=1\ll N_c$,
the back reaction from the D$7$-brane to AdS$_5\times S^5$ is negligible. 
Physically, this means that the gluon degrees of freedom is large and act as a bosonic heat bath
enabling us to have a stable nonequilibrium steady state. 
In this model we can consider quarks $\psi$ with mass $m$.
When the quarks are massive ($m\neq 0$), 
the quark and antiquark in this system are
bound to each other by a potential that are linear in short range, and
becoming Coulomb like in the long range limit~\cite{Kruczenski:2003be}. 
In the vacuum, there are no asymptotically free quarks, and the
excitation spectrum consists of a series of mesons (quark-antiquark
bound states), as well as massless gluons. 
These mesonic excitations, 
such as coherent excitations of 
$\langle\bar{\psi}\psi\rangle$, $\langle\bar{\psi}\gamma_i\psi\rangle$,
and so on, 
correspond to fluctuations for various fields
on the D$7$-brane. 
A phase transition can be induced by 
external electric fields or finite temperature 
changing the quark and antiquark excitations from 
gapped to gapless (dissociated). 
Geometrically, this corresponds to a change of the D$7$-brane embedding in the bulk spacetime
and to an emergence of an effective ``horizon''; The D$7$ fluctuations 
can now dissipate into the horizon.
The D3/D7 system and its phase transition under non-rotating electric fields, 
namely static (DC) electric fields, 
has been studied in Refs.~\cite{Karch:2007pd,Albash:2007bq,Erdmenger:2007bn,Nakamura:2012ae,Nakamura:2013yqa,Hoshino:2014nfa,Nakamura:2010zd,Hashimoto:2014yza}.

\begin{figure*}[tb]
\centering 
\includegraphics[scale=0.5]{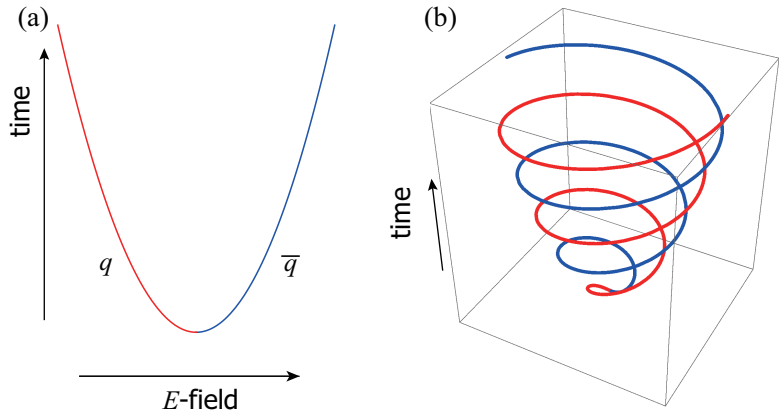}
\caption{
Schematic illustration of the DC and AC-Schwinger effects. 
(a) DC-Schwinger effect: 
In static electric fields, the quark-antiquark pair is created from the vacuum and are pulled away from each other by the field.  
(b) AC-Schwinger effect: 
In rotating electric fields, the created quark and antiquark rotate around each other as they are separated. 
}
\label{fig:Schwinger}
\end{figure*}

The electric field can make the vacuum unstable against a creation of quark-antiquark ($q$-$\bar{q}$) pairs (Fig.~\ref{fig:Schwinger}(a)), 
which is known as the Schwinger effect~\cite{Schwinger:1951} (see Refs.~\cite{Dunne2004,Gelis:review} for reviews on the QFT treatment).%
\footnote{
The holographic analogues of the Schwinger mechanism have been studied recently, 
some focusing on the elementary tunneling process described by fundamental strings 
\cite{Gorsky:2011,SemenoffZarembo,AmbjornMakeenko,Bolognesi,Sato:2013,Sato:20132,Dietrich2014,Kawai:2015},
and others studying the phase transition of the vacuum via D-brane dynamics
\cite{Hashimoto:2013,Hashimoto:2015}. 
Relations with the generation of entanglement via the EPR pairs were also discussed \cite{Sonner:2013}. 
}
The rotating electric fields can also trigger the Schwinger mechanism (Fig.~\ref{fig:Schwinger}(b)). 
In our previous work, we have studied the case when the 
quarks are massless~\cite{Hashimoto:2016ize}.
In this case, $q$-$\bar{q}$ pairs can be created without any energetic
cost and, thus, the system is always in a gapless conductive phase~\cite{Karch:2007pd} (Fig.~\ref{fig:schematic}(b)). 
A holographic version of the Floquet Weyl semimetal that shows 
anomalous quantum Hall response, i.e.,  a current can be induced that flows perpendicular to the applied electric field, 
was studied in details. 
In this paper, we holographically study the nonequilibrium steady state of strongly interacting  {\em massive quarks} in rotating electric fields. 
In the remaining of this section we summarize our findings.

\subsection{Floquet condensation and nonequilibrium phase diagram}

\begin{figure*}[bt]
\centering 
\includegraphics[scale=0.45]{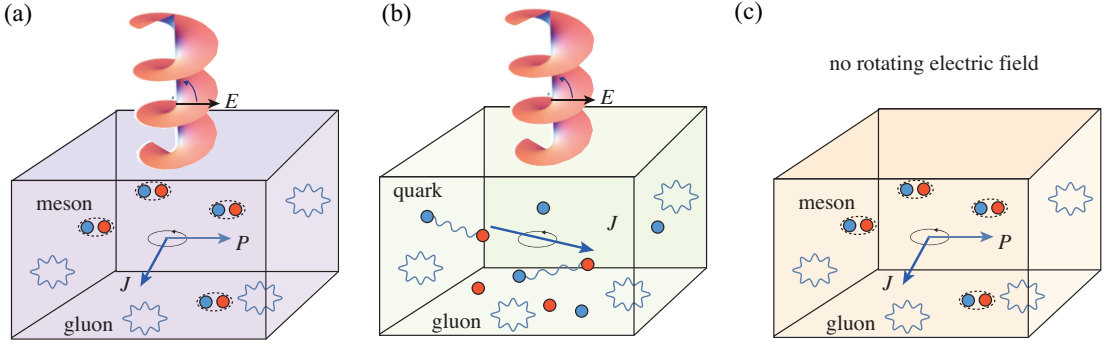}
\caption{
Schematic illustration of the states in the phase diagram Fig.~\ref{fig:3dphasediagram}.
(a) 
Insulator phase: 
Only mesons and gluons can be excited since the quark-antiquarks are gapped. 
Vacuum polarization is induced as a collective excitation of mesons
and follows the external rotating electric field. 
This induces a non-dissipative rotating current, i.e., polarization current, in the $xy$-plane with no Joule heating.  
(b) 
Conductive phase: 
Gapless excitation of quarks and antiquarks is possible and a current is induced by the external field. 
The current is dissipative with finite Joule heating.
(c) 
Vector meson Floquet condensate:  An insulator solution that has a coherently rotating current even when the external electric field is absent. 
}
\label{fig:schematic}
\end{figure*}

When quarks are massive, a single-particle excitation of the quarks or
antiquarks is forbidden and only mesons can be excited under sufficiently
weak electric fields.
It is the insulator phase of the D3/D7 system.
As illustrated in Fig.~\ref{fig:schematic}(a),
an electric field can induce ``vacuum polarization''~\cite{Schwinger:1951} by aligning the vacuum charge fluctuation. 
In  the case of QCDs, vacuum polarization is nothing but a coherent
excitation of vector mesons $\langle\bar{\psi}\gamma_i\psi\rangle$. 
In the insulator phase, by applying a rotating electric field, the polarization $\vec{\mathcal{P}}$ will follow it as
 $\vec{\mathcal{P}}\propto \vec{\mathcal{E}}$. 
The insulator phase of D3/D7 is a perfect insulator and the DC-conductivity is zero. 
However, in a time dependent field, a polarization current can exist given by the 
time derivative of the polarization
\begin{equation}
J_x+iJ_y=\bra\bar{\psi}(\gamma_x+i\gamma_y)\psi\ket= |J|e^{i\Omega t-i\delta}\ .
\label{eq:vmfc}
\end{equation}
This is perpendicular to the electric field, i.e., $\delta=\pi/2$.
Here, $\delta$ is introduced to generically parameterize the phase delay of the current relative to the electric field. 
In the insulator phase, since the current is always perpendicular to the 
electric field, Joule heating is absent and the system is dissipation-less. 
A time dependent oscillating condensate such as Eq.~(\ref{eq:vmfc}) is generically found in periodically driven systems and is called the {\em Floquet condensate} \cite{Floquetcondensate}. 
A prominent example studied theoretically in condensed matter physics is the superconducting order in 
periodically driven lattice models \cite{Murakami17, Knap16,Babadi17}.
Thus, in this article, we call the time periodic condensate expressed as Eq.~(\ref{eq:vmfc}) 
the {\em electric field induced Floquet condensate of vector mesons}. 
However, this state in the insulator phase is nothing special nor surprising; It is a state where the 
order parameter is induced by an external field. A similar example would be
a paramagnet in an oscillating magnetic field where the magnetization will  simply follow  the field. 

\begin{figure*}[tb]
\centering 
\includegraphics[scale=0.8]{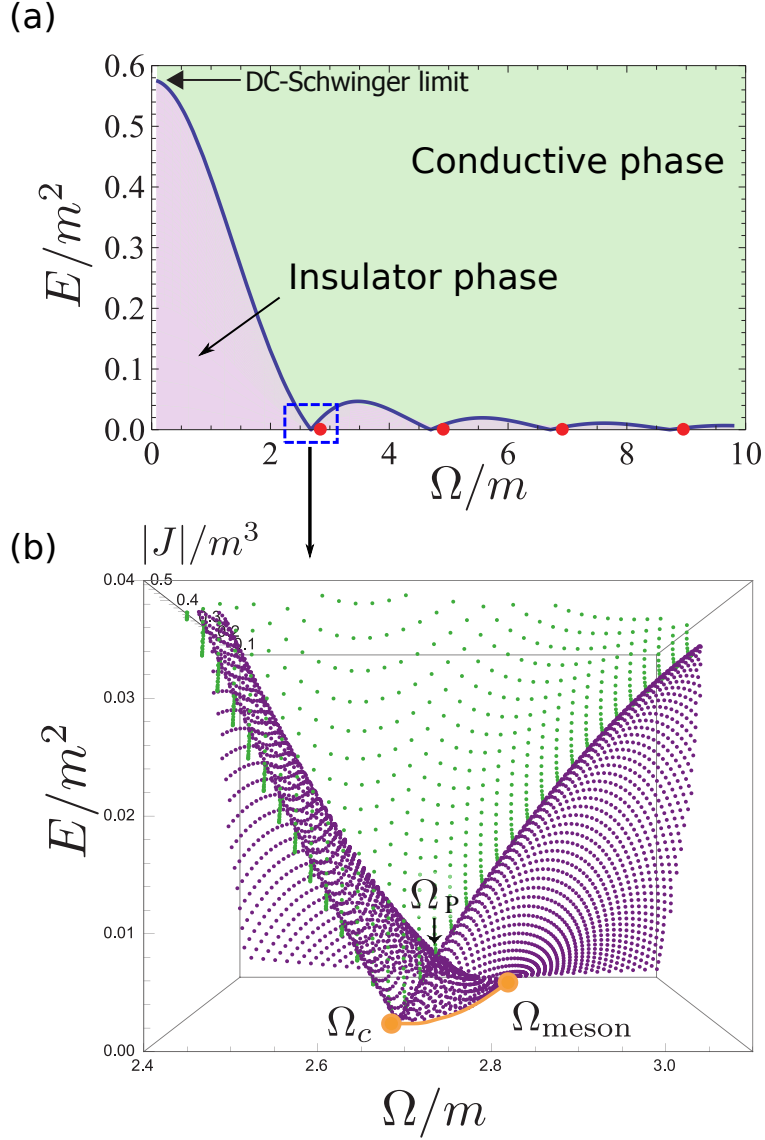}
\caption{
Main results of this paper. 
(a) The global nonequilibrium SQCD phase diagram in rotating electric fields. 
The green and purple regions represent the conductive and insulator phases, respectively. 
The critical Schwinger limit $E=E_c(\Omega)$ is shown by the solid curve, 
which becomes zero near the meson mass energy (red dots).
(b) An expanded plot of the phase diagram near the first resonance with the meson energy $\Omega_{\rm meson}$. 
The depth $|J|$ represents the amplitude of the rotating current induced by the electric field. 
The purple (green) dots denote states in the insulating (conductive) phase. 
The zero field Floquet condensate of vector mesons 
is denoted by the orange curve, an insulating solutions with a
current even at zero field $E=0$, and $\Omega_{c}$ is its end point. 
$\Omega_{\rm P}$ is the point where the two insulator-conductor boundaries cross. 
}
\label{fig:3dphasediagram}
\end{figure*}

When the field strength exceeds a threshold critical value, an insulator-to-conductor transition (i.e., dielectric breakdown) takes place. 
In the conductor phase, as illustrated in Fig.~\ref{fig:schematic}(b), gapless excitation of
quarks and antiquarks become possible. Charge carriers are now separated 
and dissipative current that are not associated to polarization can flow
allowing Joule heating to become finite. 
The critical field is known as the Schwinger limit $E=E_c(\Omega)$;
Since it is now dependent on the frequency $\Omega$, 
we coin the breakdown process induced by AC-electric fields as the  AC-Schwinger effect (Fig.~\ref{fig:Schwinger}(b)). 
The nonequilibrium phase diagram in AC-electric fields is separated into regions with 
conductive and insulating states by the AC-Schwinger limit. 
This is given in Fig.~\ref{fig:3dphasediagram} (a) 
recovering the results from section \ref{sec:phasediagram}. 
Starting from the DC case ($\Omega=0$), the critical Schwinger limit becomes smaller as the frequency $\Omega$ is increased,
and vanishes when resonance with meson modes occurs. 
Since SQCD has an infinite tower of meson mass spectrum, the threshold touches zero at each resonances, 
and thus the phase structure becomes ``lobe-shaped'' in the 
$(E/m^2,\Omega/m)$-plane.

 Figure~\ref{fig:3dphasediagram}(b) is the nonequilibrium phase diagrams focused around the first resonance.
The purple dots represent the insulating state and the green dots represent conductive states. 
Near resonance, we find an interesting state which we 
call  a zero field Floquet condensate of vector mesons, or simply a
{\em vector meson Floquet condensate}, represented by an orange line in the diagram. 
It is known that  multiple phases may coexist near the Schwinger limit \cite{Nakamura:2010zd,Nakamura:2012ae}, which 
resembles first order phase transitions in equilibrium. 
This degeneracy of states can be lifted by displaying the phase diagram as a three dimensional plot using the amplitude $|J|$ as the depth. 
At a critical frequency, which we define $\Omega_c$,  the lower boundary
of the conductive phase touches the zero field ($E=0$) plane. 
From there, on the zero field plane, a continuous set of solutions (orange line) stretches 
up to the meson mass frequency $\Omega_{\rm meson}$. 
The vector meson Floquet condensate, as illustrated in Fig.~\ref{fig:schematic}(c),
is a subset of the insulator phase solutions, which
supports a rotating current $J$ even when the external driving field is zero. 
The current breaks time translational symmetry 
as well as the time reversal symmetry ($\Omega\to -\Omega$). 
This state exists not only for the lowest meson resonance but for all resonances in the vector meson mass spectrum. 
It  is a state with a coherent oscillation of ``vacuum polarization''~\cite{Schwinger:1951}.

The vector meson Floquet condensate is expected to be dynamically stable within the present theory 
since mesonic excitations cannot dissipate in the probe limit ($N_f=1 \ll N_c$).
If the system is isolated from an energy bath, there is no relaxation channel within the model that changes this state, in other words,
it is a fixed point of time evolution, i.e., a
non-thermal fixed point \cite{Berges:2004}.
We note the relation of the Floquet condensate discussed here with more exotic states known as ``time crystals''. 
The ground state time crystal proposed by 
F. Wilczek is, in our language, a Floquet condensate that is thermodynamically stable
existing as a lowest energy state  \cite{Wilczek:2012}. 
Another class of time crystal is the ``Floquet time crystal''~\cite{Khemani:2016,Else:2016}, a state 
in a periodically driven closed quantum system showing oscillations with a
periodicity longer than the external driving force that was experimentally studied in
artificial matters \cite{ZhangTC:2016,ChoiTC:2016}. 
The vector meson Floquet condensate is not included in neither of these categories
since it is an excited state, and the periodicity is the same as the driving.

\subsection{Possible experiments: Bose Einstein condensation of chiral excitons in gapped Dirac materials}

\begin{figure*}[htb]
\centering 
\includegraphics[scale=0.4]{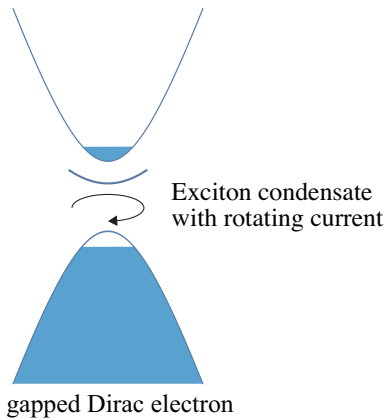}
\caption{Chiral exciton BEC in gapped Dirac materials: 
This figure schematically shows the weak coupling picture of the vector meson Floquet condensate. 
It is equivalent to the chiral exciton BEC that can be realized in gapped Dirac materials such as bulk topological insulators. }
\label{fig:weakcoupling}
\end{figure*}

The mesons in the current D3/D7 system  shows
many similarities with excitons discussed in solid state physics. 
Quark-antiquark pairs and the interaction mediated by gluons
replace the electron-hole pairs and Coulomb interaction there. 
The vector meson Floquet condensate is then the counterpart of an exciton Bose Einstein condensate (BEC). 
Exciton BEC is already experimentally realized in semiconductors \cite{Yoshioka:2011}. 
However, in contrast to the non-relativistic electrons in semiconductors,
the quarks and antiquarks in the D3/D7 system are Dirac particles that have spin and angular momentum degrees of freedom. 
Thus, if we translate the states in our theory to the condensed matter language, 
it should be excitons in gapped Dirac materials in the strong coupling
limit, i.e., BEC limit (Fig.~\ref{fig:weakcoupling}). 
It is possible to find materials that host gapped 3D Dirac electrons: bulk
Bismuth \cite{Bismuthreview}, and three dimensional topological insulators \cite{TI1,TI2} such 
as Bi$_2$Se$_3$ and its family materials are good candidates. 
In such materials, by applying circularly polarized laser, 
one can excite excitons with an orbital angular momentum.  The chirality, i.e., 
the sign of the angular momentum, is selected by the helicity of the laser. 
The vector meson Floquet condensate in SQCD is 
translated to the {\em Bose Einstein condensate (BEC) of chiral excitons} in gapped Dirac materials. 
We note that related theoretical models which shows oscillating exciton condensation are currently being studied 
\cite{Murakami_exciton18,Tanay18}.

The formation of Floquet states and its quasi-energy spectrum were 
observed in the surface two dimensional Dirac band a time resolved angle-resolved photoemission 
spectroscopy (ARPES) experiment done in a three dimensional topological insulator \cite{Gedik1}. 
This was done with circularly polarized laser whose photon energy is lower than the bulk gap. 
In order to realize BEC of chiral excitons, a laser with photon energy close to the bulk gap should be used. 
If BEC is realized, our theory predicts that
even after the pump laser pulse has disappeared
a Floquet state will remain that is maintained by the bulk rotating current.

\begin{figure*}[tbh]
\centering 
\includegraphics[scale=0.8]{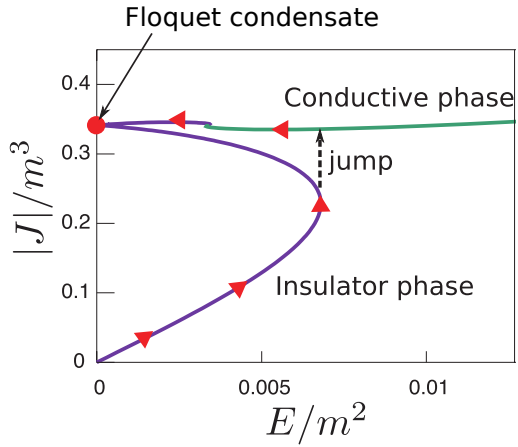}
\caption{
The amplitude of the current as a function of $E$ for a fixed $\Omega \in (\Omega_c,\Omega_{\rm meson})$
discussed in more detail in Sec.~\ref{sec:physical_quantities}.
Arrows show a presumable process to realize the vector meson Floquet condensate (chiral exciton BEC).
}
\label{fig:realize_bec}
\end{figure*}

How can we efficiently create a chiral exciton BEC with a large rotating current? 
The phase diagram Fig.~\ref{fig:3dphasediagram} gives us some hints. 
First, using a pulse field, it is possible to directly create the BEC
if the peak strength is large enough.  This process is illustrated in Fig.~\ref{fig:realize_bec},
which is a constant $\Omega$-slice of the phase diagram Fig.~\ref{fig:3dphasediagram}. 
Imagine that the field is given by 
$\mathcal{E}_x+i\mathcal{E}_y= E(t) e^{i\Omega t}$
where the envelope $E(t)$ changes slowly. 
In a pulse field, $E(t)$ increases from zero up to a maximum and then decrease. 
If this process is slow enough, the state will adiabatically follow  the insulator state, 
as indicated by red arrows, until the state no longer exists. 
Then, a abrupt jump will take place to the other solution, which is conductive for $\Omega_c<\Omega<\Omega_{\rm P}$ 
and insulating for $\Omega_{\rm P}<\Omega<\Omega_{\rm meson}$. 
As the field weakens, the state will follow the solution down 
to the chiral exciton BEC. 
This process is similar a hysteresis loop of magnetization in ferromagnets.

Second, the chirping technique, i.e., time dependent shift of frequency $\mathcal{E}_x+i\mathcal{E}_y= E e^{i\Omega (t) t}$
may be useful in following the chiral BEC solution up to large rotating current. 
As Figure~\ref{fig:3dphasediagram}(b) suggests, 
the resonant frequency of the condensate (orange line) redshifts from $\Omega_{\rm meson}$ to $\Omega_{c}$
as the condensation density increases. 
We can adiabatically follow the orange line by using a chirped pulse in which the laser frequency $\Omega(t)$
slowly redshifts. The merit of this 
approach is that there is no threshold in the field strength and can be done with relatively weak fields. 
There is an upper limit in the magnitude $|J|$ of the rotating current in the 
insulator phase since a insulator to conductive transition takes place
at $\Omega=\Omega_c$. 
This transition is nothing but the exciton (meson) Mott transition 
\cite{Mott:1968,Zimmermann:1988,Suzuki:2012,He:mesomMott,Hashimoto:2014yza}
which happens when the exciton density exceeds a threshold
at which the Coulomb attraction that binds the excitons becomes strongly screened and 
the excitons dissolve into plasma of electrons and holes.

Another important realization of a rotating electric field is heavy ion collision. 
Calculations suggest that strong electromagnetic fields can be created when heavy ions fly by each other~\cite{Kharzeev:2007jp,Skokov:2009qp,Voronyuk:2011jd,Bzdak:2011yy,Deng:2012pc}.
The field that is created in such processes mimics a half-cycle rotating electric field. 
If the frequency of the pulse is  resonant with the meson excitation energy, 
it is possible to produce vector meson Floquet condensates by such fields efficiently.

The rest of the paper is organized as follows. 
In Sec.~\ref{Dbranetime},
we will see that the equations of motion for the D$7$-brane can be
reduced to ordinary differential equations effectively 
by moving to the co-rotating frame~\cite{Hashimoto:2016ize}, and in Sec.~\ref{sec:embeddings}
we will explain three types of their solutions with different brane embeddings.
On the basis of solutions obtained numerically, we show a nonequilibrium
phase diagram and physical quantities of the dual gauge theory near the
phase boundary in Secs.~\ref{sec:phasediagram} and 
\ref{sec:physical_quantities}. 
In Sec.~\ref{sec:high_frequency}, we will give an analytical derivation
of some features in 
the phase diagram for the high frequency limit.
In Sec.~\ref{sec:toy}, we provide a simple understanding of the 
phase diagram using a toy model.


\section{Formalism for a D-brane in external rotating electric fields}
\label{Dbranetime}

\subsection{Time-dependent setup for D-brane}

We consider $\mathcal{N}=2$ supersymmetric QCD as the field theory.
Its gravity dual is described by the dynamics of a probe D7-brane in the AdS$_5\times S_5$ spacetime.
The metric of $\mathrm{AdS}_5 \times S^5$ is written as
\begin{equation}
 ds^2=\frac{\rho^2+w_1^2+w_2^2}{R^2}[-dt^2+dx^2+dy^2+dz^2]
  +\frac{R^2}{\rho^2+w_1^2+w_2^2}[d\rho^2 + \rho^2 d\Omega_3^2 + dw_1^2 +dw_2^2]\ ,
  \label{AdS5S5}
\end{equation}
where $R$ is the AdS radius, $\rho=\infty$ is the AdS boundary. Hereafter, we take the unit $R=1$.
We study the D7-brane in this geometry. 
Dynamics of the D7-brane is described by the Dirac-Born-Infeld (DBI) action,
\begin{equation}
 S=-T_7 \int d^8\sigma \sqrt{-\textrm{det}[h_{ab}+2\pi\alpha' F_{ab}]}\ ,
\end{equation}
where $T_7$ is the tension of the brane,
$h_{ab}$ is the induced metric and 
$F_{ab}=\partial_a A_b-\partial_bA_a$ is the $U(1)$-gauge field strength 
on the brane worldvolume.
In the AdS$_5\times S^5$ spacetime, the $(t,x,y,z,\rho,\Omega_3)$-directions are filled with the D7-brane.
We assume spherical symmetry in $S^3$ and translational symmetry in the $(x,y,z)$-space.
We also set $A_t=A_\rho=A_z=0$ for simplicity, which means zero baryon
number density in the boundary theory. We write the worldvolume gauge field as
\begin{equation}
 (2\pi \alpha')  A_a d\sigma^a = a_x(t,\rho) dx + a_y(t,\rho) dy\ .
  \label{arealdef}
\end{equation}
For later convenience, we introduce a complex expression for the gauge field,
\begin{equation}
 a(t,\rho)\equiv a_x(t,\rho)+ia_y(t,\rho)\ .
  \label{coma}
\end{equation}
For the brane position, we can consistently assume%
\footnote{
Because of the rotational symmetry in the $(w_1,w_2)$-plane, 
the schematic form of the equation of motion for $\vec{w}=(w_1,w_2)$ becomes
$\ddot{\vec{w}}=f_1 \vec{w}''+f_2 \vec{w}' + f_3 \vec{w}$,
where $f_{1,2,3}$ are invariant under $U(1)$-rotation.
It follows that $w_2=0$ is a consistent solution.
}
\begin{equation}
 w_1=w(t,\rho)\ ,\qquad w_2=0\ .
\end{equation}
Although we take into account the full time-dependence of the gauge fields  and the brane position,
we will see that we can eliminate the time-dependence by a change of variables in next subsection.
In terms of the brane embedding $w$ and the complex field $a$, 
the DBI action is written as
\begin{multline}
S= -T_7 \Omega_3 V_3 \int dt d\rho 
 \frac{\rho^3}{(w^2+\rho^2)}\bigg\{
 (1+w'{}^2)(\rho^2+w^2)^2
 +[(w^2+\rho^2)^2|a'|^2-\dot{w}^2]\\
 -]
\dot{w}^2|a'|^2+(1+w'{}^2)|\dot{a}|^2-2\dot{w}w'\textrm{Re}(\dot{a}a'{}^\ast)
]
-[\textrm{Im}(\dot{a}a'{}^\ast)]^2
\bigg\}^{1/2}\ ,
\label{action_a}
\end{multline}
where $\Omega_3=\textrm{Vol}(S^3)$ and $V_3=\int dxdydz$.
From the equations of motion derived from the above action,
we obtain an asymptotic solution near the AdS boundary ($\rho=\infty$) as
\begin{align}
 a(t,\rho)&=-\int^t dt' \mathcal{E}(t') + \frac{\mathcal{J}(t)}{2\rho^2} 
 +\frac{\dot{\mathcal{E}}(t)}{2\rho^2} \ln \left(\frac{\rho}{\rho_0}\right) + \cdots\ ,\label{ainf}\\
w(t,\rho)&= m +\frac{c}{\rho^2} +\cdots\ ,
\label{awinf}
\end{align}
where $\rho_0$ is a constant with the dimension of length 
and the dot denotes a $t$-derivative.
$\mathcal{E}(t), \mathcal{J}(t)\in \mathbb{C}$ and $m, c\in \mathbb{R}$
are expansion coefficients, which depend on time in general.%
\footnote{
For the rotating electric field, we will see that $w(t,\rho)$ can be time-independent.
Therefore, we have dropped the argument $t$ for $m$ and $c$.
}
They are related to the electric field $\mathcal{E}_\textrm{SQCD}(t)$,
electric current $\mathcal{J}_\textrm{SQCD}(t)$, quark mass $m_q$
and quark condensate $\langle \mathcal{O}_m\rangle$ in the boundary theory by
\begin{equation}
\begin{split}
&\mathcal{E}_\textrm{SQCD}(t)=\left(\frac{\lambda}{2\pi^2}\right)^{1/2}\mathcal{E}(t)\ ,\quad
\mathcal{J}_\textrm{SQCD}(t)  = \frac{N_c\sqrt{\lambda}}{2^{5/2}\pi^3}\, \mathcal{J}(t)\ ,\\
&m_q=\left(\frac{\lambda}{2\pi^2}\right)m\ ,\quad
 \langle \mathcal{O}_m\rangle = -\frac{N_c\sqrt{\lambda}}{2^{3/2}\pi^3}\, c\ .
 \label{winf}
\end{split}
\end{equation}
Note that we use the complex notation for the electric field and current:
real and imaginary parts of $\mathcal{E}$ and $\mathcal{J}$ correspond to
$x$- and $y$-components of the electric field and current,
such as $\mathcal{E} = \mathcal{E}_x + i\mathcal{E}_y$ and 
$\mathcal{J} = \mathcal{J}_x + i\mathcal{J}_y$.

\subsection{Co-rotating frame and reduction to an 1-dimensional problem}
\label{subsec:reduction}

We focus on the rotating external electric field that is described by 
\begin{equation}
\mathcal{E}(t)= E e^{i\Omega t}\ ,
  \label{rotEbc}
\end{equation}
where $E$ is a complex constant whose argument represents the initial
direction of the electric field at $t=0$.
The electric field rotates anti-clockwise with a frequency $\Omega$ 
in the $(x,y)$-plane. 
From Eq.~(\ref{ainf}),
the boundary condition for the gauge field $a$ at the AdS boundary $\rho=\infty$ becomes
\begin{equation}
 a|_{\rho=\infty}=\frac{iE}{\Omega} e^{i\Omega t}\ .
  \label{bc_a}
\end{equation}
We define two real variables $b(t,\rho)$ and $\chi(t,\rho)$ as 
\begin{equation}
a(t,\rho)=b(t,\rho)e^{i [\Omega t + \chi(t,\rho)]}\ ,
\label{bdef}
\end{equation}
where $b$ and $\chi$ represent the amplitude and the phase for $a$,
respectively.
For new variables $b$ and $\chi$, the boundary condition becomes time-independent as
\begin{equation}
 be^{i\chi}|_{\rho=\infty}=\frac{E}{\Omega}\ .
  \label{bc_b}
\end{equation}
Note that, since $\chi$ at the AdS boundary 
describes the direction of
the electric field at a certain time, one can fix it to an arbitrary value.

Using new variables $b(t,\rho)$ and $\chi(t,\rho)$, the DBI action~(\ref{action_a}) can be rewritten as
\begin{multline}
S= -T_7 \Omega_3 V_3 \int dt d\rho 
 \frac{\rho^3}{(w^2+\rho^2)}\bigg\{
 (\rho^2+w^2)^2(1+w'{}^2+b'^2) + b^2(w^2+\rho^2)^2\chi'^2
 \\
 - [1+(b'^2 + b^2\chi'^2)]\dot{w}^2
 - (1+w'^2)|\dot{b}+i(\Omega + \dot{\chi})b|^2
 \\
 + 2w' \dot{w} [b'\dot{b} + b^2\chi'(\Omega+\dot{\chi})]
 - b^2[\dot{b}\chi' - (\Omega + \dot{\chi})b']^2
 \bigg\}^{1/2}\ .
\label{action_b}
\end{multline}
This action does not depend on $t$ explicitly.
The boundary condition~(\ref{bc_b}) is also time-independent.
Thus, we can consistently assume that the variables $b$, $\chi$ and $w$
do not depend on $t$:
$b(t,\rho)=b(\rho)$, $\chi(t,\rho) = \chi(\rho)$ and $w(t,\rho)=w(\rho)$.
By the reduction to 1D problem, the DBI action~(\ref{action_b}) is simply written as
\begin{equation}
 \begin{split}
&S= -T_7 \Omega_3 V_4 \int d\rho \mathcal{L}_0\ ,\\
&\mathcal{L}_0\equiv \frac{\rho^3}{(w^2+\rho^2)}
\sqrt{
[(\rho^2 + w^2)^2 - \Omega^2 b^2](1+w'^2+b'^2)
 + (\rho^2 + w^2)^2 b^2 \chi'^2
  } \ ,
 \end{split}
\label{DBI_static}
\end{equation}
where $V_4=\int dtdxdydz$.
Obviously, this action is invariant under the constant shift of the phase variable $\chi(\rho)$: 
$\chi(\rho) \to \chi(\rho) + \alpha$, 
where $\alpha$ is an arbitrary real constant.
As mentioned above, this arbitrariness of a constant phase rotation of $a$
corresponds to changing the initial direction of the electric field in the $(x,y)$-plane in the
boundary theory.
Thus, we have a conserved charge defined by  
\begin{equation}
 q \equiv \Omega\frac{\partial\mathcal{L}_0}{\partial \chi'}
  = \frac{\Omega \rho^6}{\mathcal{L}_0} b^2 \chi' \ .
\label{qjule}
\end{equation}
As we will see later, the above conserved charge $q$ corresponds to the Joule
heating in the boundary theory.
This implies that $\Omega \chi'(\rho) \ge 0$ should be satisfied to make the
Joule heating $q$ non-negative.
The ansatz for the gauge field in Eq.~(\ref{bdef}) is similar to that of an
ordinary circularly polarized electromagnetic wave in the Maxwell theory.
In fact, $\Omega \chi'(\rho) \ge 0$ is nothing but the condition for
the ``circularly polarized'' gauge field to propagate inward along
the $\rho$-direction from the AdS boundary on the D$7$-brane.
The above conserved quantity $q$ is the energy flux of this
``circularly polarized'' gauge field 
in the bulk theory.

From Eq.~(\ref{DBI_static}), equations of motion for $b$, $\chi$ and
$w$ are given by%
\footnote{
For numerical calculations,
$b(\rho)$ and $\chi(\rho)$ are not suitable variables
when a solution approaches the origin of the complex plane: $b(\rho)=0$ although they are convenient for the simple expressions. 
(Note that Eq.~(\ref{chieq}) is singular at $b(\rho)=0$
because the phase $\chi(\rho)$ becomes ill-defined
.)
To avoid this ``coordinate singularity'',
we use 
Cartesian-like variables 
$B_x(\rho) + i B_y(\rho) \equiv b(\rho)e^{i\chi(\rho)}$
in our actual numerical calculations.
We can easily cast the equations of motion or boundary conditions for
$(b(\rho),\chi(\rho))$ into those for 
$(B_x(\rho),B_y(\rho))$.
}
\begin{align}
 b'' =& \frac{1}{\rho (w^2+\rho^2)[(w^2+\rho^2)^2 - \Omega^2b^2]}
 \Big\{
 - 3(w^2+\rho^2)^3 b' (1+w'^2+b'^2)
  \nonumber\\
 & 
 - \Omega^2b (1+w'^2+b'^2)
 \left[\rho(w^2+\rho^2) -  bb' (3w^2+\rho^2)\right]
 \nonumber\\
 & 
 + b (w^2+\rho^2)^3 (\rho - 3bb')\chi'^2
 \Big\} \ ,
 \label{beq}
 \\
 \chi'' =& \frac{1}{b\rho (w^2+\rho^2)[(w^2+\rho^2)^2 - \Omega^2b^2]}
 \Big\{-3 b^3(w^2+\rho^2)^3 \chi'^3
 \nonumber\\
 & - 
 \left[
 2\rho b'(w^2+\rho^2)^3 + 3b(w^2+\rho^2)^3(1+w'^2+b'^2)
 \right.
 \nonumber\\
 &
 \left. -  \Omega^2 b^3
 \left(4\rho^2 + 4\rho ww' + (\rho^2 +3w^2)(1+w'^2+b'^2)\right)
 \right]\chi'
 \Big\} \ ,
 \label{chieq}
 \\
 w''=&\frac{1}{\rho(w^2+\rho^2)\{(w^2+\rho^2)^2-\Omega^2 b^2\}}\Big\{
 -3(w^2+\rho^2)^3(1+w'{}^2+b'^2)w'\nonumber\\
 &
 + \Omega^2 b^2
 \left[2\rho w + (\rho^2+3w^2)w'\right](1+w'^2+b'^2)
 \nonumber\\
 &
 - 3 b^2 (w^2+\rho^2)^3 w' \chi'^2
 \Big\}\ .
 \label{weq}
\end{align}
It turns out that $\chi(\rho) = \text{const.}$ is a trivial solution of
Eq.~(\ref{chieq}).
In this case, the conserved quantity $q$ is zero and the gauge
field becomes a standing wave without a net energy flux rather than a
propagating wave.
As we will see later, whether such solutions can be admitted physically
depends on boundary conditions for the other fields $b(\rho)$ and
$w(\rho)$.

It is worth noting that we can easily recover the DC limit from the expression of
the action (\ref{DBI_static}).
This is done by fixing 
$\Omega b(\rho) \to E_i$ and $\chi'(\rho)/\Omega \to a_i'(\rho)/E_i$ while taking  the $\Omega\to 0$ limit,
where $i$ denotes a direction determined by the alignment at $t=0$, 
and then we recover the expression for static electric fields \cite{Hashimoto:2013}. 
This is because this system still holds a translational invariance with
respect to $\partial_t + \Omega(x\partial_y - y\partial_x)$ instead of
time translational invariance with respect to $\partial_t$.
Therefore, the rotating electric field we are studying now is a natural
generalization of static electric fields.
Since the vector $\partial_t + \Omega(x\partial_y - y\partial_x)$
represents a co-rotating frame with angular velocity $\Omega$, the
above translational invariance implies that the system should be stationary in
the co-rotating frame.
This enables us to treat the system in a similar way to static
systems and provides its phase structure sharply as we will see later.

\subsection{Observables in the boundary quantum field theory}
\label{obsbdry}

Once we solve the equations of motion,
observables in the boundary theory can be read out from asymptotic
behavior of the solution near the AdS boundary $\rho=\infty$.
From the asymptotic form of $w(\rho)$, we can read off quark mass and condensate
as in Eq.~(\ref{winf}). Multiplying $e^{-i\Omega t}$ to Eq.~(\ref{ainf}) and using the rotating electric field ansatz~(\ref{rotEbc}),
we obtain the asymptotic form of $b(\rho)e^{i\chi(\rho)}$ as
\begin{equation}
 b(\rho)e^{i\chi(\rho)}
 =\frac{iE}{\Omega}+\frac{J}{2\rho^2}+\frac{i\Omega
  E}{2\rho^2}\ln \left(\frac{\rho}{\rho_0}\right) +\cdots\ .
\label{binf}
\end{equation}
We have defined $J$ by
\begin{equation}
 \mathcal{J}(t)=Je^{i\Omega t}\ ,
\end{equation}
where $J\in \mathbb{C}$ is a time-independent constant 
representing the electric current at $t=0$.
In particular, we have $E = -i\Omega be^{i\chi}|_{\rho=\infty}$ from Eq.~(\ref{binf}).  
By using the $U(1)$-symmetry of selecting the initial field direction, i.e., shift of $\chi|_{\rho=\infty}$, 
we can fix $E$ to be a real and non-negative value.

Substituting the above expression into Eq.~(\ref{qjule}), which is the
conserved charge in the bulk, we can rewrite the conserved charge $q$ 
as 
\begin{equation}
 q=\text{Re}(E^\ast J)\ .
\label{qinf}
\end{equation}
This  explicitly confirms that the conserved quantity $q$ in the
bulk theory corresponds to the Joule heating in the boundary theory.
Besides the expression (\ref{binf}), the complex electric current can be written as 
\begin{equation}
 J = \lim_{\rho\to\infty}
  \left[\rho^3\left(b'+ib\chi'\right)
   - \frac{\Omega^2b}{2}\ln\left(\frac{\rho^2}{e\rho_0^2}\right)
		    \right]e^{i\chi} ,
  \label{jasy}
\end{equation}
where the last term in the bracket is arising from a counter-term in holographic
renormalization~\cite{Karch:2007pd}.
It is known that the electric current has an ambiguity corresponding to a
finite counter-term if the electric field is time-dependent, and 
in fact the ambiguity is reflected in a constant $\rho_0$ of (\ref{jasy}).
However, it is worth noting that the Joule heating (\ref{qinf}) is unaffected by
this ambiguity as is clear from (\ref{jasy}). 
In general, an electric current in a dielectric medium can be decomposed
into the {\em active current} contributing to the Joule heating via resistive
dissipation and the {\em reactive current} (or polarization current) originating in polarizations. 
It turns out that a part of $J$ proportional to $\chi'(\rho)$ represents
the active current, while the remnant is the reactive current.

There is a scaling symmetry in our setup: 
\begin{equation}
\begin{split}
 &t\to t/\alpha\ ,\quad \rho \to \alpha \rho\ ,\quad w\to \alpha w\ ,
 \quad
 b\to \alpha b\ ,\quad
 \chi\to \chi\ .\\
&\Omega\to \alpha \Omega\ ,\quad
 E\to \alpha^2 E\ ,\quad
 J\to \alpha^3 J\ ,\\
&m\to \alpha m\ ,\quad
 c\to \alpha^3 c\ ,\quad
q\to \alpha^5 q\ .
\end{split}
\label{eq:scaling}
\end{equation}
We will nondimensionalize physical quantities using the quark mass $m$.

\section{Black hole, Minkowski and critical embeddings}
\label{sec:embeddings}

The equations of motion~(\ref{beq})-(\ref{weq}) are singular
at $\rho=0$ and $\rho=\rho_c$ where $\rho_c$ satisfies
\begin{equation}
 b(\rho_c)= \frac{w(\rho_c)^2+\rho_c^2}{\Omega} \ .
\label{brhoc}
\end{equation}
The singular points $\rho=0$ or $\rho=\rho_c$ give the inner boundary of the integration of equations of motion.
In section~\ref{sec:eff}, we will see that
$\rho=\rho_c$ is the effective horizon with respect to the effective
metric on the D$7$-brane.
There are three possibilities of the brane solutions depending on the position of the inner boundary:
\begin{enumerate}
 \item The brane reaches $\rho=0$, the axis of $S^3$ wrapped by the
       D$7$-brane, and the effective horizon does not exist.
       (i.e. no $\rho_c$ satisfying Eq.~(\ref{brhoc}) exists.)
       \label{enum:Minkowski}
 \item The brane intersects with the effective horizon at $\rho=\rho_c$
       before it reaches the axis.
       \label{enum:BH}
 \item The brane reaches the axis which coincides with the locus of the
       effective horizon, namely $\rho_c =0$.
       \label{enum:critical}
\end{enumerate}
The brane embeddings in the cases~\ref{enum:Minkowski}, \ref{enum:BH}
and \ref{enum:critical} are called
``Minkowski embeddings'', ``black hole embeddings'' and ``critical
embeddings'', respectively~\cite{Frolov:2006tc,Mateos:2006nu,Mateos:2007vn}.

\begin{figure}
 \centering
 \includegraphics[scale=0.6]{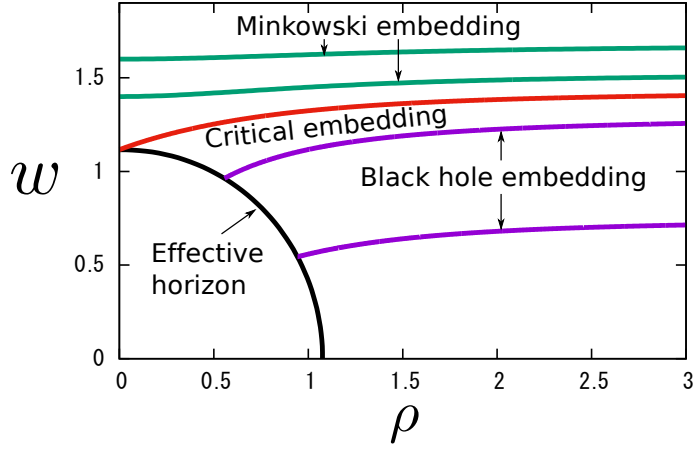}
 \caption{
 Typical profiles of the D$7$-brane in the ($\rho,w$)-plane 
 for $E=1$ and $\Omega=1$.
 }
 \label{E1Om1emb}
\end{figure}

Figure~\ref{E1Om1emb} shows typical profiles of the D$7$-brane for the black
hole, Minkowski and critical embeddings in the $(\rho,w)$-plane.
We fix the electric field and its angular velocity as $E=1$ and $\Omega=1$.
Asymptotic position of the brane corresponds to the quark mass. (See Eq.~(\ref{winf}).)
The trajectory of the effective horizon emerging on the D$7$-brane for fixed $E=\Omega=1$
is shown by the black curve in the bulk spacetime. (Note that there is no horizon in the bulk spacetime.)
For static electric fields $\Omega=0$, it is known that the trajectory of
the effective horizon is given by a circle: $w^2+\rho^2=E$ for a fixed $E$~\cite{Albash:2007bq,Erdmenger:2007bn,Hashimoto:2014yza}.
In this figure,
by the effect of non-zero $\Omega$,
the trajectory expands and is slightly deformed from the round circle.

\subsection{Minkowski embeddings}

Here, we consider the Minkowski embeddings (case~\ref{enum:Minkowski}).
In this case the conserved quantity of (\ref{qjule}) leads to $q=0$ because
the brane reaches $\rho=0$.
Furthermore, since $q=0$ should be conserved for $\rho > 0$, we have the
trivial solution $\chi(\rho) = \text{const.}$
We will focus on $b(\rho)$ and $w(\rho)$, and expand the solutions
around $\rho=0$.
The asymptotic solutions become
\begin{align}
&b(\rho)=b_0-\frac{\Omega^2 b_0}{8(w_0^4-\Omega^2 b_0^2)}\rho^2+\cdots\ ,\label{baxis}\\
&w(\rho)=w_0+\frac{\Omega^2 b_0^2}{4w_0(w_0^4-\Omega^2 b_0^2)}\rho^2+\cdots\ ,\label{waxis}
\end{align}
where $w_0^2-\Omega b_0> 0$.
When the brane does not intersect with the effective horizon, no
Joule heating is generated in the boundary theory.
As we mentioned, $\chi(\rho)=\text{const.}$ means the worldvolume gauge
field becomes standing wave without any net energy flux.

\subsection{Black hole embeddings}

  In the case of the black-hole embeddings (case~\ref{enum:BH}) there
  exists $\rho_c (>0)$ satisfying (\ref{brhoc}).
  The effective horizon emerges at $\rho=\rho_c$ on the D$7$-brane.
  Equations (\ref{qjule}) and (\ref{brhoc}) directly give us the
  conserved quantity as 
  \begin{equation}
   q = \Omega\rho_c^3 b(\rho_c)
    = \rho_c^3[w(\rho_c)^2 + \rho_c^2] .
    \label{q_horizon}
  \end{equation}
  Thus, in contrast to the Minkowski embeddings $\chi(\rho)$ becomes
  non-trivial and the energy flux appears on the D$7$-brane.
  In the boundary theory this means Joule heating $q\neq 0$ occurs due to the current. 
  
  The boundary conditions for $w(\rho)$, $b(\rho)$ and $\chi(\rho)$ are
  given by the regularity at the effective horizon $\rho=\rho_c$.
  Expanding the solutions around $\rho=\rho_c$, we obtain
  \begin{equation}
   \begin{aligned}
    w(\rho) =& w_0 + w_1 (\rho - \rho_c) + \cdots\
    ,\\
    b(\rho) =& b_0 + b_1 (\rho - \rho_c) + \cdots\
    ,\\
    \chi(\rho) =& \chi_0 + \chi_1 (\rho - \rho_c) + \cdots\ , 
   \end{aligned}
  \end{equation}
  where $b_0 = (w_0^2 + \rho_c^2)/\Omega$ and $\chi_0$ is arbitrary.
  Explicit forms of the first derivatives $w_1$, $b_1$ and $\chi_1$ are
  shown in appendix \ref{app:BC}.

\subsection{Critical embeddings}

Solutions of the phase boundary between the Minkowski embeddings and the
black-hole embeddings are the critical embeddings (case \ref{enum:critical}).
In this case the locus of the effective horizon coincides with the axis,  
so that (\ref{brhoc}) yields $b_0 = \frac{w_0^2}{\Omega}$ because of
$\rho_c=0$.
We expand the solutions around $\rho=0$ as 
\begin{align}
 &b(\rho)=\frac{w_0^2}{\Omega}
 - \frac{\Omega}{\sqrt{24w_0^2 + 6\Omega^2}} \rho
 +\cdots\ ,\label{baxis_c}\\
 &w(\rho)=w_0
 + \frac{2w_0}{\sqrt{24w_0^2 + 6 \Omega^2}} \rho
 +\cdots\ .\label{waxis_c}
\end{align}
Since $q=0$, solution of $\chi(\rho)$ is arbitrary constant.

\subsection{Effective metric and horizon}
\label{sec:eff}

The effective metric is defined by~\cite{Seiberg:1999vs,Gibbons:2000xe,Gibbons:2001ck,Gibbons:2002tv,Kim:2011qh}
\begin{equation}
 \gamma_{ab}=h_{ab}+(2\pi\alpha')^2 F_{ac}F_{bd}h^{cd}\ .
\end{equation}
For the current system, the effective metric is written as
\begin{multline}
 \gamma_{ab}d\sigma^a d\sigma^b
 =-F(\rho)dt^2+\frac{2\Omega b^2 \chi'}{w^2+\rho^2}dtd\rho
 +\frac{(1+w'{}^2+b'^2)}{w^2+\rho^2}d\rho^2\\
 +\frac{(w^2+\rho^2)}{4(1+w'{}^2)}
 [(b'+i\chi'b)e^{i\chi}e_- + (b'-i\chi'b)e^{-i\chi} e_+]^2
+F(\rho)e_+ e_-
 \\
 +\frac{\Omega^2b}{4(w^2+\rho^2)}(e^{i\chi}e_- + e^{-i\chi}e_+)^2
 +(w^2+\rho^2)dz^2+\frac{\rho^2}{w^2+\rho^2}d\Omega_3^2\ ,
\end{multline}
where
\begin{equation}
 F(\rho)=\frac{(w^2+\rho^2)^2-\Omega^2 b^2}{(w^2+\rho^2)}\ .
\end{equation}
We have also defined 1-forms $e_\pm$ as
\begin{equation}
 e_\pm=e^{\pm i\Omega t}(dx\pm idy)\ .
\end{equation}
Note that $F(\rho)$ becomes zero at $\rho=\rho_c$. This implies that $\rho=\rho_c$ is the event horizon with respective to the brane effective metric $\gamma_{ab}$.
In general, dynamics of the brane itself and various fields on the brane is
governed by the effective (induced) metric.
The effective horizon is a causal boundary for fluctuations on the
brane, so that the fluctuations across the effective horizon can never
go back and will dissipate.

The effective Hawking temperature is given by
\begin{equation}
 T_H=\left.\frac{-\gamma_{tt}'}{4\pi\gamma_{t\rho}}\right|_{\rho=\rho_c}
  = \left.\frac{2(ww'+\rho_c)-\Omega b'}{2\pi b\chi'}
    \right|_{\rho=\rho_c}\ ,
\end{equation}
where explicit expressions of $w'$, $b'$ and $\chi'$ at the effective
horizon $\rho=\rho_c$ are shown in appendix \ref{app:BC}.

\section{Nonequilibrium SQCD phase diagram in rotating electric fields}
\label{sec:phasediagram}

Here we explain the structure of the nonequilibrium phase diagram 
that we obtain by solving the equation of motions (\ref{beq})-(\ref{weq}) for the probe D7-brane. 
There are two phases. 
The insulator phase is given by the Minkowski embedding solutions.  
In this phase fluctuations on the D$7$-brane, correspond
to mesonic excitations, can never dissipate, so that it is a gapped
phase where stable mesons can exist.
Since we have $\chi(\rho)=\text{const.}$,
the Joule heating is absent although the rotating
current $J$ is finite.
This means that this current is a reactive current with no resistive
dissipation.
The conductive phase is given by the black hole embeddings.
Fluctuations will dissipate into the effective horizon
emerging on the D$7$-brane. 
This phase is gapless where the mesons have been dissociated and 
there are no stable mesons.
Now, $\chi(\rho)$ becomes a non-trivial  function and an active current
with the Joule heating emerges.
This phase emerges after the dielectric breakdown of the insulator phase.
The Joule heating plays a role as an order parameter to distinguish the phase transition
between insulator and  conductive phases.

We should describe the phase diagram in a physical parameter space. 
The brane solutions are originally specified by the boundary conditions
with three parameters $\Omega$, $b_0$, and $w_0$ at the inner boundary; 
for the Minkowski embeddings we specify $b_0$ and $w_0$
($\Omega b_0 < w_0^2$) at the axis $\rho=0$, and for the black hole
embedding $b_0$ and $w_0$ at the effective horizon
$\rho=\rho_c$, where $\rho_c = \sqrt{\Omega b_0 - w_0^2}$. 
Using the scaling symmetry (\ref{eq:scaling}), we can set one of the parameters to unity without loss of generality
\footnote{In our numerical calculation, we set $\Omega=1$.}.
This means that the phase diagram of our system can be described by a
two-parameters family of solutions, specified by the frequency $\Omega/m$ and strength $E/m^2$ of the rotating electric field.

\begin{figure}
  \centering
  \includegraphics[scale=0.8]{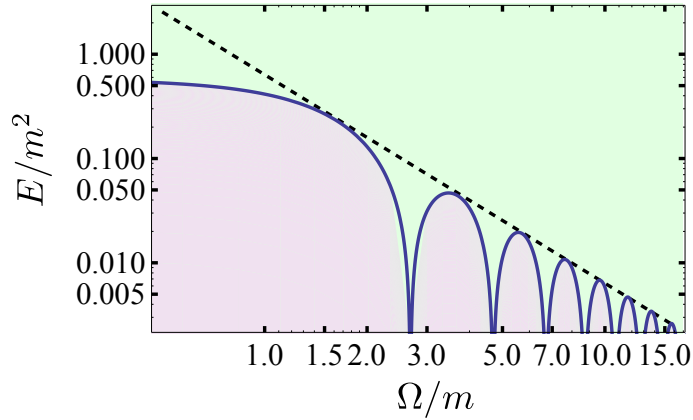}
  \caption{
  Frequency dependence of the AC-Schwinger limit shown on a log-log plot.
  The solid curve represents the AC-Schwinger limit.
  The dashed line is $E/m^2 = \frac{2}{\pi}(\Omega/m)^{-2}$.
  }
  \label{fig:loglog_phase}
 \end{figure}

Figure~\ref{fig:3dphasediagram}(a) 
shows a phase diagram in the
parameter space $(\Omega/m, E/m^2)$.
The phase boundary between the insulator and the conductive phases is given by the 
critical embeddings (solid curve) giving the AC extension of the
Schwinger limit for dielectric
breakdown~\cite{Schwinger:1951,Heisenberg1936,Weisskopf}, 
that is, the AC-Schwinger limit. 
The region above the curve indicates the conductive phase
while that below indicates the insulator phase, basically.
(As will see, actually, multiple phases co-exist in the
vicinity of the critical curve.)
We see a lobe-shaped phase structure, 
i.e., the insulator phase is separated into infinite but discrete regions analogous to a lobe (leaf).
We can see that ``lobes'' become thinner as the frequency becomes larger.
Its height decreases following a power-law $\Omega^{-2}$
(Fig.~\ref{fig:loglog_phase}),
which can be analytically verified in 
Sec.~\ref{sec:high_frequency}.
The red points on the horizontal axis are mass spectrum of mesons in $\mathcal{N}=2$ SQCD:  
$\Omega_\textrm{meson}/m = 2\sqrt{(n+1)(n+2)}$ $(n=0,1,2,\ldots)$~\cite{Kruczenski:2003be}.
The critical embedding solution (solid curve) touches the horizontal
axis at discrete points smaller than but close to the meson mass frequencies.
We refer to this frequency as the critical frequency and denote it as $\Omega_c$.
The numerical values of the critical frequencies and meson mass spectrum are given by
\begin{equation}
\begin{split}
 \Omega_c/m&=2.6828,\ 4.6985,\ 6.7147,\ 8.7275,\cdots\ ,\\
 \Omega_\textrm{meson}/m&=2.8284,\ 4.8990,\ 6.9282,\ 8.9443, \cdots\ ,
\end{split}
\end{equation}
for the first four modes. 
In Sec.~\ref{sec:high_frequency}, we will examine more deeply how the
critical  and meson mass frequencies appear by employing a perturbative approach
in a high-frequency regime.
We find that the AC-Schwinger limit periodically becomes
infinitesimal near, but not equal to, frequencies corresponding to the
meson mass spectrum.
This indicates that 
the rotating electric fields can resonantly excite the vector mesons
close to the meson mass frequencies, so the dielectric
breakdown can easily occurs even though the rotating
electric field is too weak.

We have overviewed the global structure of the phase diagram obtained from the critical solution.
Next, let us focus on the detailed structure of the phase diagram near the phase boundary.
Discrete self-similarity is known to emerge near the phase boundary, 
so that multiple
Minkowski and black hole embedding solutions  with a spiral structure take place near the phase boundary
resulting in a characteristic first order phase transition. 
Such behavior 
has been studied in various probe brane systems~\cite{Frolov:2006tc,Mateos:2006nu,Mateos:2007vn}.
The self-similarity leads to oscillatory behavior of physical quantities 
as one approaches the transition. 
Figure~\ref{fig:3dphasediagram}(b) 
shows
fine structure of the phase diagram around the lowest meson excitation energy $\Omega_\mathrm{meson}/m=2\sqrt{2} \simeq 2.83$.
We take $|J|$ as the third axis to distinguish degenerated states for a same ($\Omega/m, E/m^2$).
We plot Minkowski and black hole embedding solutions 
by purple and green dots that are obtained by the changing parameter set ($w_0,b_0$)%
\footnote{
For the Minkowski embedding, we set parameters at inner boundary as
$\Omega=1$, $w_0=1/\alpha$ and $b_0=(w_0^2/\Omega)\beta$, where 
$\alpha=2.0+0.01i$ and $\beta=0.01j$ ($i=0,1,\cdots,150$, $j=0,1,\cdots 96$).
For the black hole embedding, we set
$\Omega=1$, $w_0=(1/\alpha)\sin \beta$ and $\rho_c=(1/\alpha)\cos \beta$, where 
$\alpha=2.0+0.02i$ and $\beta=0.01j$ ($i=0,1,\cdots,70$, $j=0,1,\cdots,56$).
}.
We see that both solutions co-exist for the same parameters 
($\Omega/m, E/m^2$) near the critical electric field. 
Such multiplicity of the solutions was observed in the
case of the static electric fields ($\Omega=0$), so that the above
feature is a natural AC extension of that around the Schwinger limit. 
It is especially notable that
the multiplicity extends to the horizontal axis ($E=0$)
within a range of the frequency: 
$\Omega_c\leq \Omega \leq \Omega_\textrm{meson}$
(see Fig.~\ref{fig:3dphasediagram} also). 
Among the two states at $E=0$, one is the thermal ground state, while the other is 
a non-trivial excited state.  
This non-trivial state in the absence of external electric fields can be
interpreted as a non-linear excitation of the vector mesons: 
Floquet condensate of the vector mesons.
It is known that the vector mesons can stably exist in the insulator
phase.
In general, coherent excitations of these vector mesons appear as 
oscillations of the expectation values 
with frequencies corresponding to each masses.
However, if the vector mesons are circularly polarized, its amplitude
does not oscillate while its phase oscillates (the direction of the
polarization is rotating).
When the excitation is small enough to be described by linear
perturbations, the frequency is given by the meson mass 
$\Omega_\mathrm{meson}$.
As the amplitude becomes larger, nonlinearity of the excitation cannot
be negligible and its frequency will shift from the meson mass frequency.

When multiple states co-exist, in the long time limit, a state that has lesser energy will be 
selected as a thermodynamical stable state, while the other 
may exist as a transient meta-stable state. 
Within the present model, although  
Floquet condensate of the vector mesons 
is thermodynamically unstable, 
this meta-stable state is expected to be a dynamically stable regarding time-evolution
and defines a ``non-thermal fixed point''. 
This is because the mesonic excitations can be stable
without dissipating in the insulator phase.

If we fine-tune the frequency of the rotating electric field $\Omega$,
we can control the insulator-to-conductor phase transition by a weak electric field.
To see this explicitly, let us focus on $\Omega=\Omega_P$ in Fig.~\ref{fig:3dphasediagram}(b).
At this point, we have $(\Omega_\textrm{P}/m, E_\textrm{P}/m^2)=(2.728,5.292\times 10^{-3})$.
Above this point ($E>E_\textrm{P}$, $\Omega=\Omega_\textrm{P}$), only the conductive phase exist.
This indicates that, for this frequency $\Omega=\Omega_\textrm{P}$,
we can cause the phase transition just by $E/m^2=E_\textrm{P}/m^2=5.292\times 10^{-3}$ at most. 
Note that, for DC electric field $\Omega=0$,
we need $E/m^2\simeq 0.6$ for the phase transition. (See the vertical axis in Fig.~\ref{fig:3dphasediagram}(a).)
This is 100 times larger than the case of the fine-tuned frequency $\Omega=\Omega_\textrm{P}$.

 \section{Physical quantities near the phase boundary}
 \label{sec:physical_quantities}

 In order to understand the meaning of the multiple solutions for given 
 ($\Omega/m, E/m^2$), we focus on their physical quantities in the dual
 gauge theory.
 From the asymptotic brane solution near the AdS boundary,
 we can read off two vacuum expectation values in the dual gauge theory.
 One of them is the rotating electric current $J$, which appears as the
 non-linear response against the external rotating electric field.
 Note that $J$ is a complex number, whose amplitude and phase represent the
 magnitude and the initial direction of the electric current, respectively.
 The other is the quark condensate $c$.
 As the subsequent quantity, we also consider the Joule heating 
 $q=\mathrm{Re}(E^\ast J)$
 playing the role of the order parameter of the phase transition 
 between the insulator and conductive phases.
 The Joule heating becomes non-zero only in the conductive phase,
 while the electric current can exist in the both conductive and
 insulator phase.
 It is a contrast to the case of the static
 electric fields ($\Omega=0$) in which the insulator phase has neither the
 electric current nor the Joule heating.

 \begin{figure}
  \centering
  \subfigure[$\Omega\leq\Omega_c=2.6828m$]
  {\includegraphics[scale=0.5]{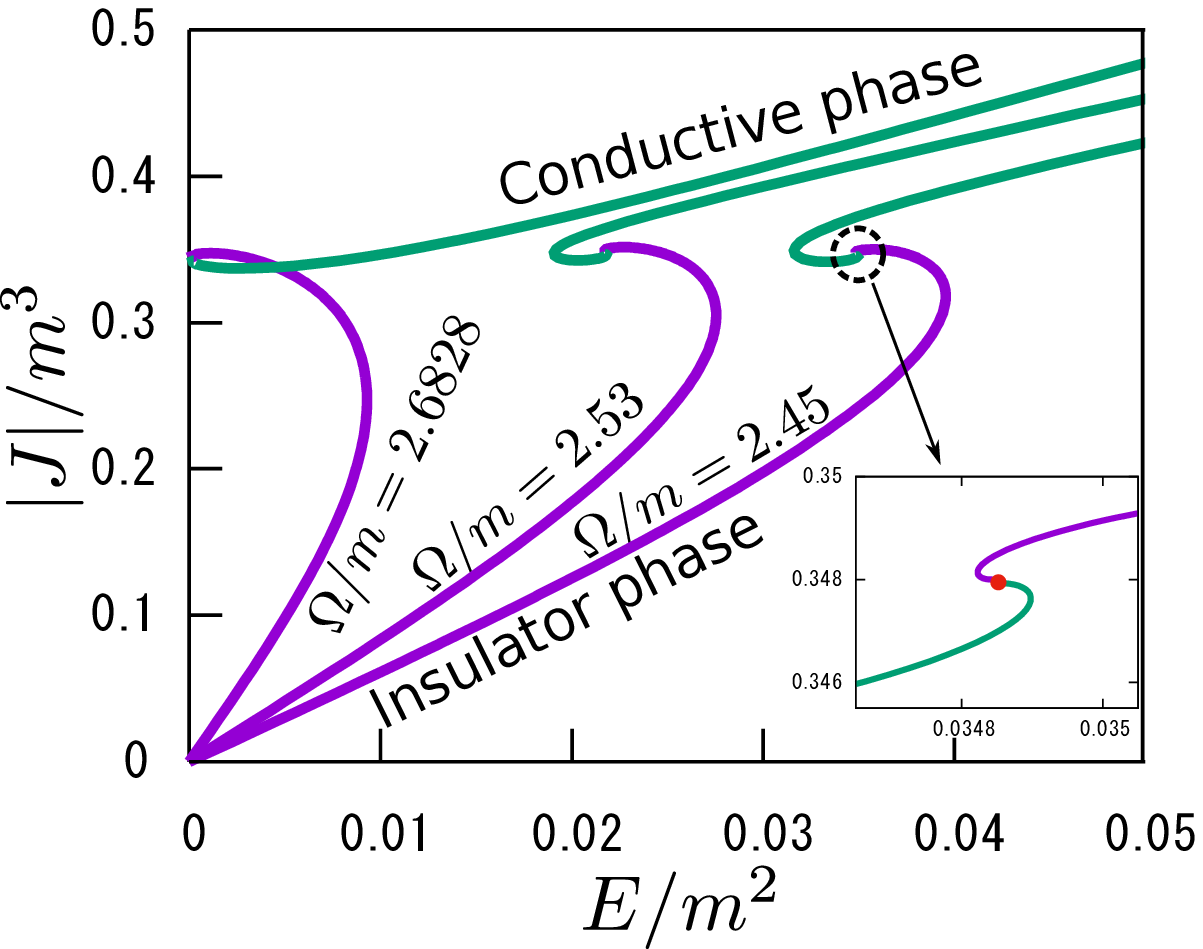}\label{JEbef}
  }
  \subfigure[$\Omega>\Omega_c=2.6828m$]
 {\includegraphics[scale=0.5]{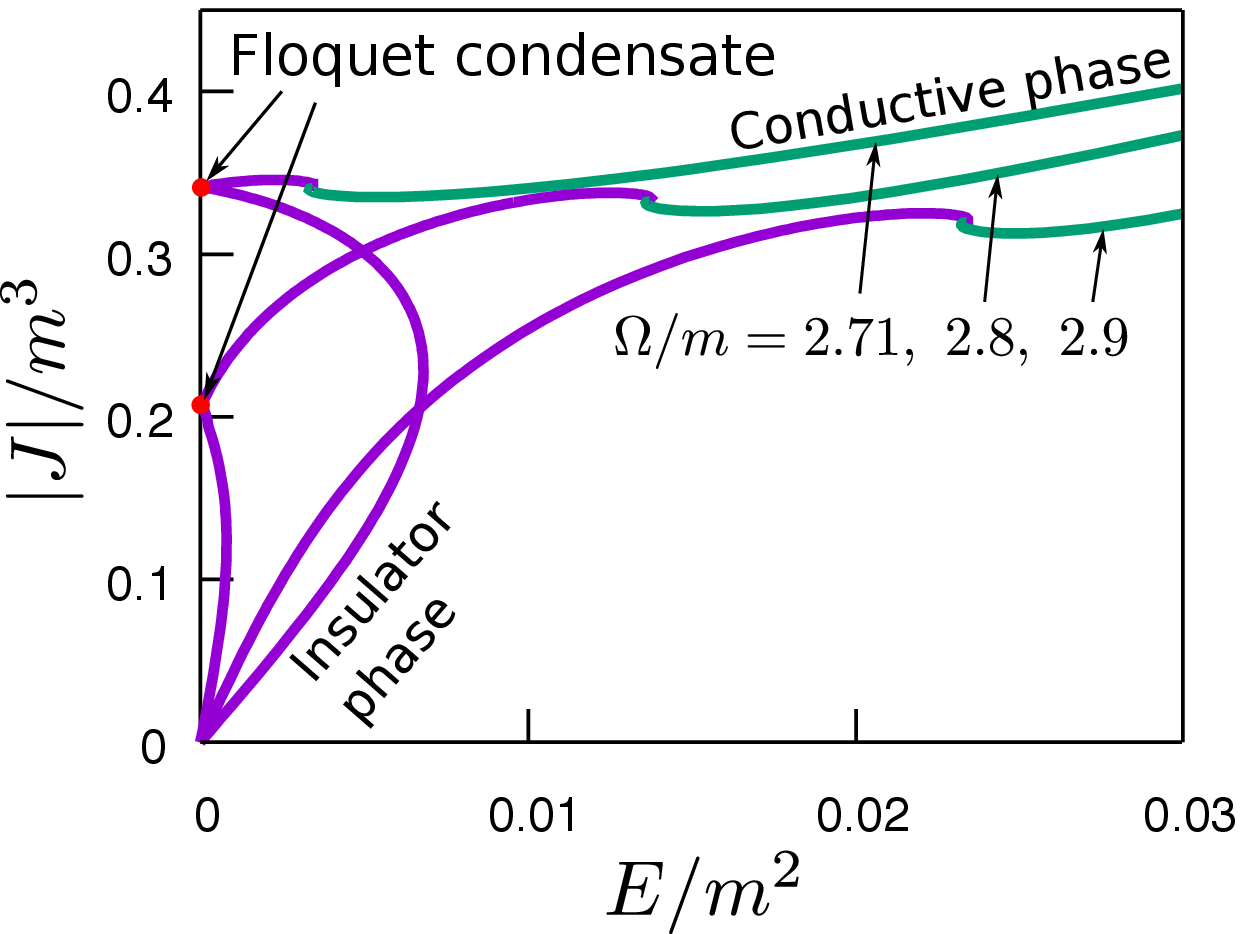}\label{JEaf}
  }
  \caption{
  Absolute values of magnitude of the rotating electric currents as functions of the electric field for fixed several $\Omega$.
  The left (right) panel shows data for frequencies below (above)  the  critical frequency.
  The purple and green curves are for insulator and conductive phases, respectively.
  The inset of the left panel shows the spiral structure at the phase boundary.
  }
 \label{E_vs_current}
 \end{figure}

\begin{figure}
  \centering
  \subfigure[$\Omega\leq\Omega_c=2.6828m$]
  {\includegraphics[scale=0.45]{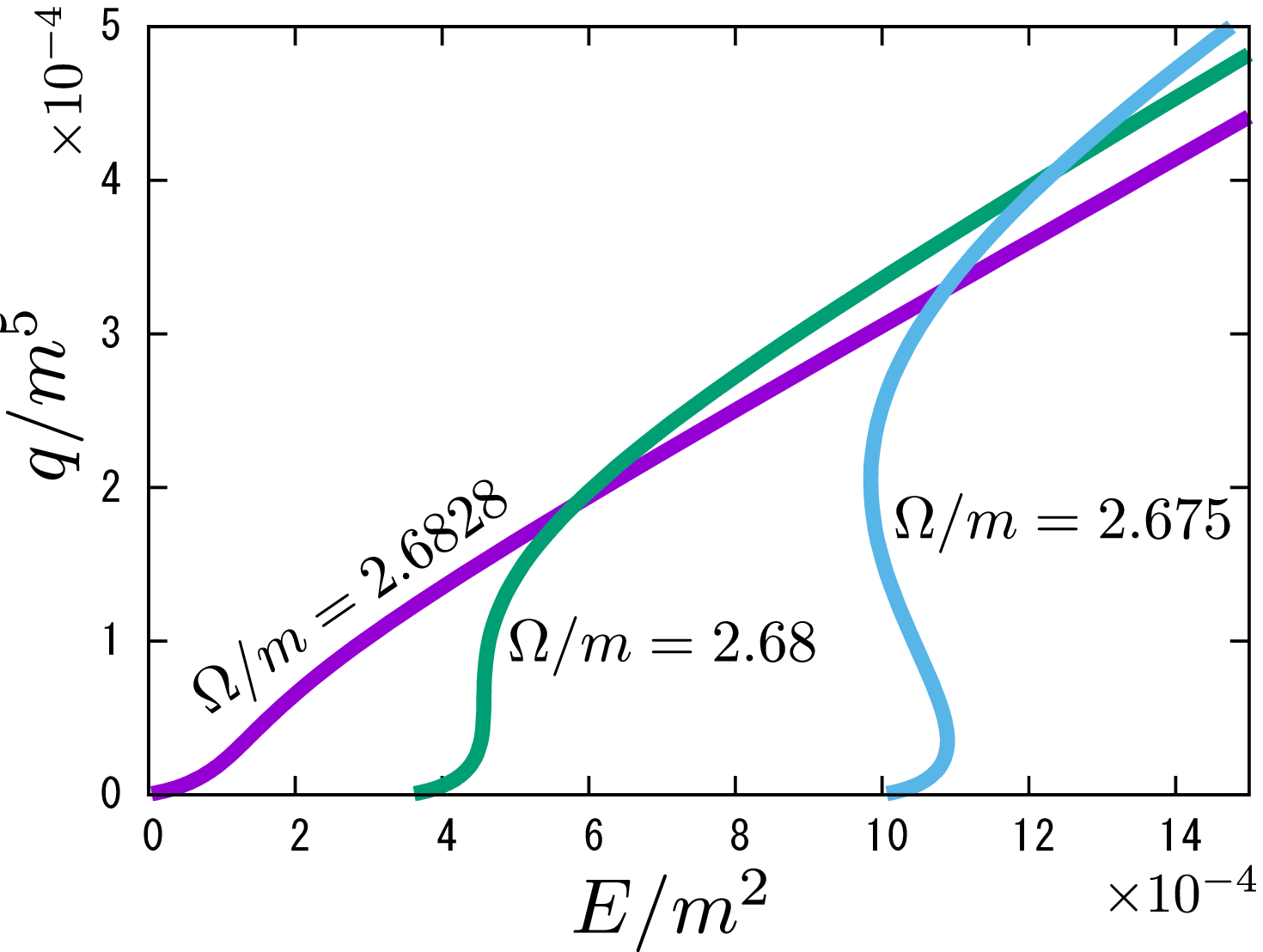}\label{JouleEbef}
  }
  \subfigure[$\Omega\geq \Omega_c=2.6828m$]
 {\includegraphics[scale=0.45]{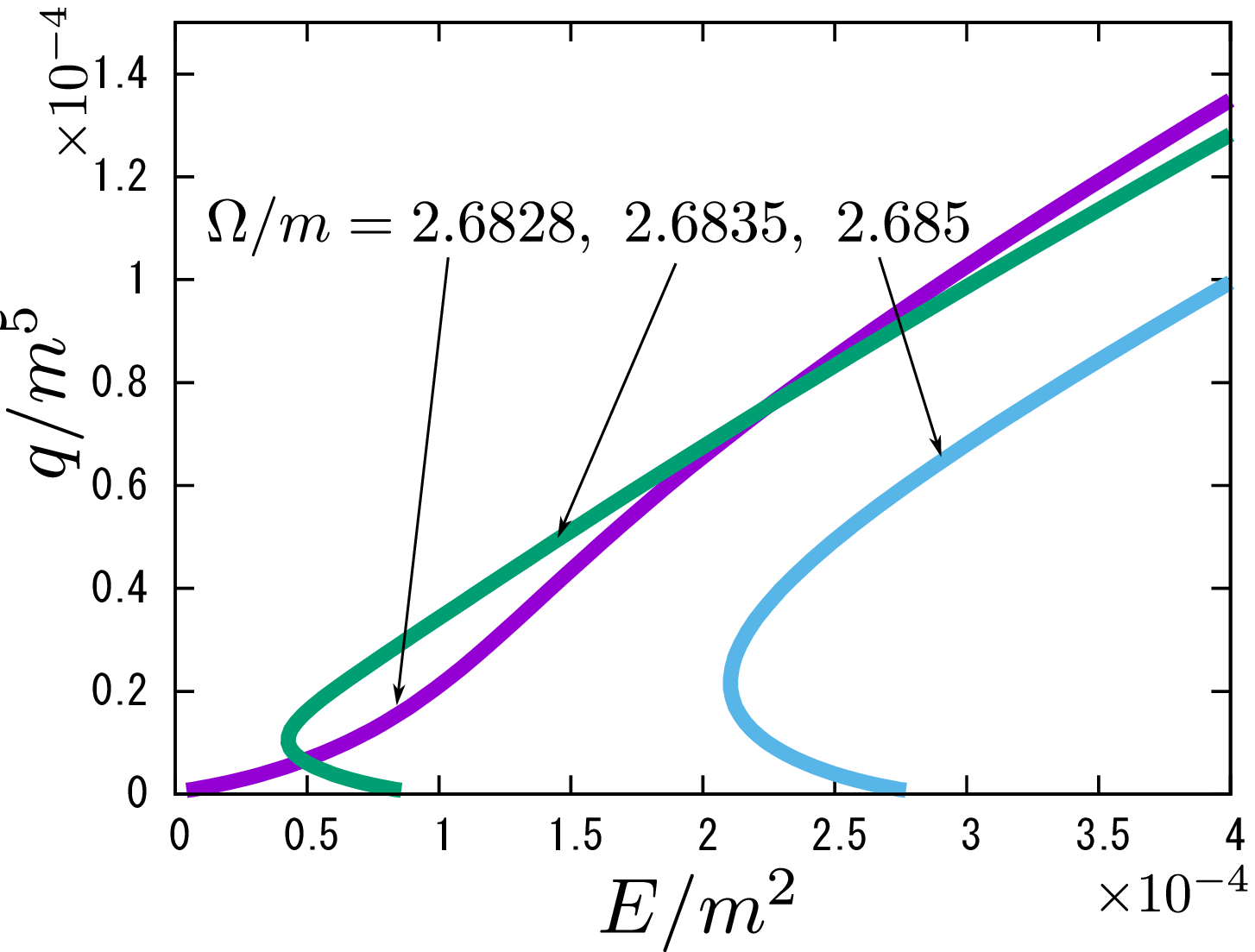}\label{JouleEaf}
  }
  \caption{
  The Joule heating $q/m^5$ is depicted as
 functions of the electric field $E/m^2$ for some fixed $\Omega/m$. Only conductive phases are shown.
 The left and right panels are for below and above critical frequency.
 }
 \label{fig:E_vs_Joule}
\end{figure}

Figure~\ref{E_vs_current} shows the electric currents as functions of 
the electric field $E$ for several fixed frequencies $\Omega$.
Near the phase boundary, the electric current becomes multivalued.
In the multivalued region, the most stable state is expected to be realized. 
If we start increasing the external field from $E=0$, the system goes through an insulator-to-conductor transition at the critical electric field.
Then, there will be a jump in the electric current as well in the Joule heating at the phase transition.
(There is also a jump in the Joule heating. See Fig.~\ref{fig:E_vs_Joule}.)
From this perspective, the phase transition from insulator to conductor is ``first order''.
Such phase transitions between nonequilibrium states with the electric
current have been often observed in a similar kind of holographic
systems described by D3/D7-branes, especially around the critical
embedding solutions.

Spiral structure of the solutions can be seen taking place 
around the critical embedding solution as shown in the  inset of the left panel of Fig.~\ref{E_vs_current}. 
As we increase the frequency
from $\Omega<\Omega_c$ to $\Omega=\Omega_c$ (left panel),
the center of the spiral approaches the vertical axis $E=0$
and touches at the critical frequency $\Omega=\Omega_c$.
Between $\Omega_c<\Omega<\Omega_\textrm{meson}$ (right panel),
the insulator phase continues to touch and reflect at the vertical axis $E=0$.
At reflecting points, the electric field is zero but the current is non-zero. 
This is 
the Floquet condensate of the vector mesons, 
which is the non-thermal fixed points solutions (existing between $\Omega_c<\Omega<\Omega_\textrm{meson}$). 
As explained in section~\ref{obsbdry}, the definition of the electric
current contains an ambiguity with a term proportional to the electric field.
(The ambiguous parameter $\rho_0$ in Eq.~(\ref{jasy}) is set to $\rho_0=m$.)
However, only the current at $E=0$ is determined without this ambiguity. 
This means that the expectation value of the Floquet condensate obtained
here is definite at $E=0$. 

 \begin{figure}
  \centering
  \subfigure[$\Omega\leq\Omega_c=2.6828m$]
  {\includegraphics[scale=0.5]{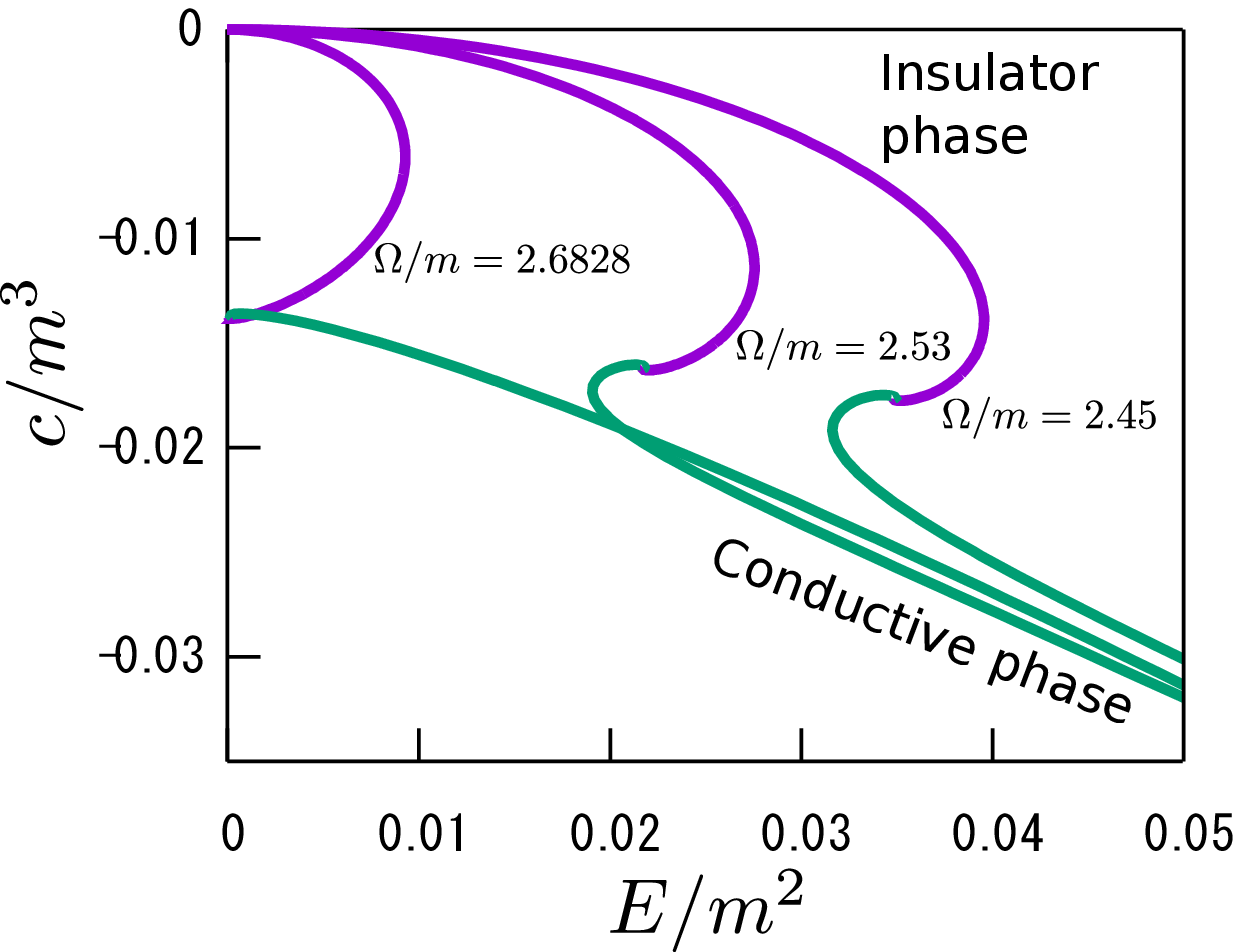}\label{cEbef}
  }
  \subfigure[$\Omega>\Omega_c=2.6828m$]
 {\includegraphics[scale=0.5]{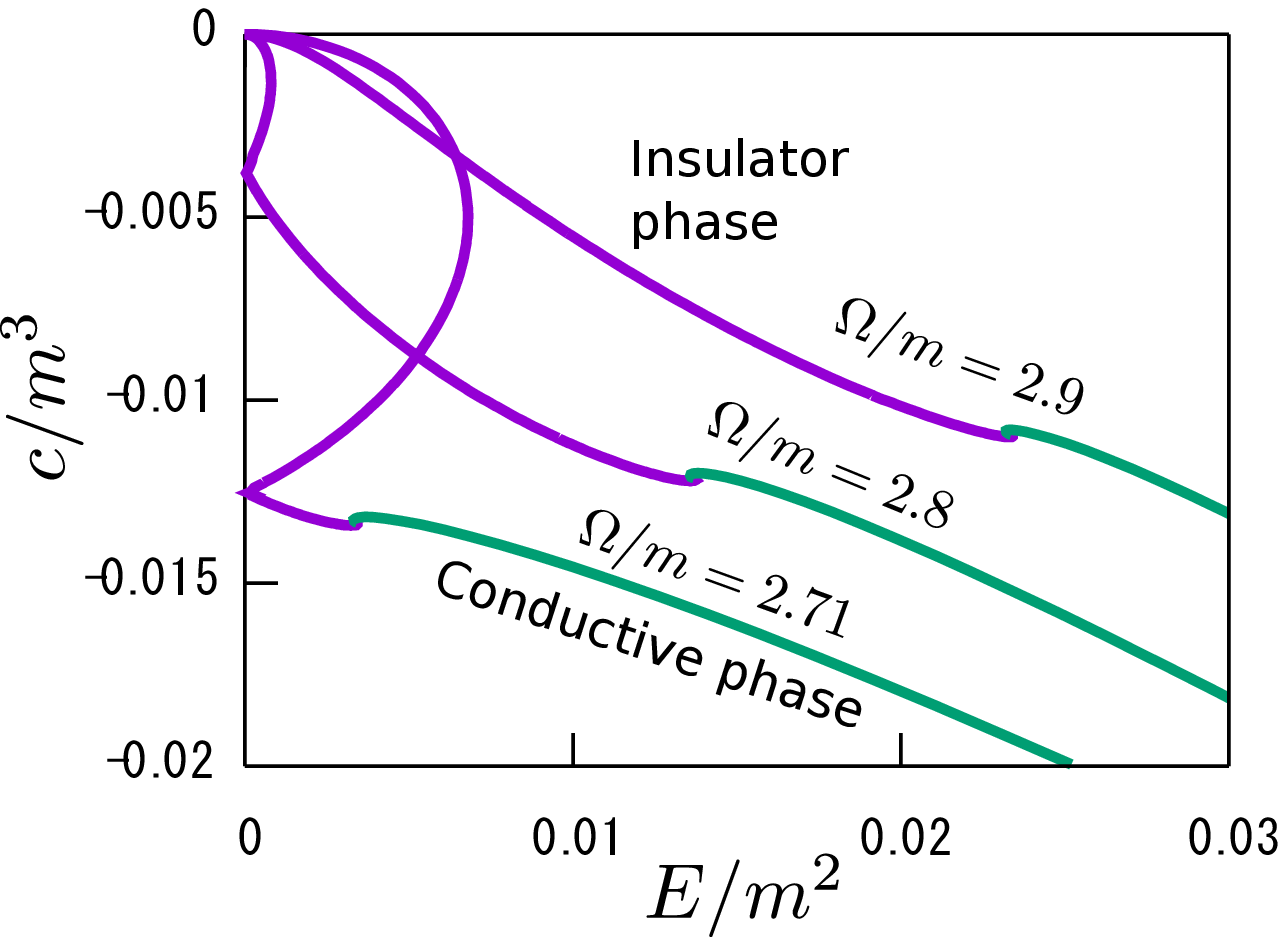}\label{cEaf}
  }
  \caption{
  Quark condensates as functions of the electric field for fixed several $\Omega$.
  The left and right panels are for below and above critical frequency.
  The purple and green curves are for insulator and conductive phases, respectively.
  }
 \label{E_vs_c}
 \end{figure}

Figure~\ref{E_vs_c} shows the quark condensates as functions of 
the electric field $E$ for several fixed frequencies $\Omega$.
Note that, for the definition of the quark condensate, there is no ambiguity as in the electric current.
The basic behavior of the condensate is very similar to that of the electric current~Fig.~\ref{E_vs_current}
and we find it quite remarkable that the quark condensate is also
non-zero 
as well as the Floquet condensate of the vector mesons.

Compared with  meson condensate realized by linearly polarized fields 
($\mathcal{E}_x\propto \cos\Omega t$ and $\mathcal{E}_y=0$),
the condensate realized by rotating electric field is distinct as follows.
The rotating electric field breaks time translational and reversal
symmetries and makes the system a time periodic steady state, i.e.,
a Floquet state.
Indeed its vector quantities such as the directions of the electric
current and the polarization oscillate, but 
its scalar quantities such as the condensation of a scalar meson, and
the magnitude of the polarization associated with vector mesons, do not
oscillate.

 Figure~\ref{fig:E_vs_Joule} shows the Joule heating $q$ as functions of
 the electric field $E$.
 This quantity is only nonzero in the conductive phase. 
 We see that the Joule heating is also multi-valued in terms of 
 the electric field. 
 It is remarkable that, within a small region of the frequency $2.68 \lesssim \Omega/m \lesssim 2.6828$, the
 multiplicity of the Joule heating
 disappears
 and the nonequilibrium phase transition between conductive phases 
 does not occur~\cite{Nakamura:2012ae}. 
 In general, the multiplicity around the critical solutions may be infinitely large, i.e., 
 infinite number of solutions can coexist for the same value of the electric field. 
 In the current numerical analysis, it is still unclear whether all of the infinite multiplicity
can disappear
around the critical frequency.

 \begin{figure}
 \centering
 \includegraphics[scale=0.5]{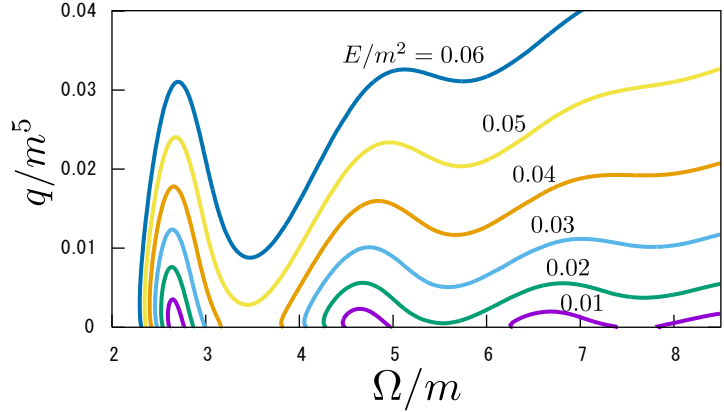}
  \caption{
 Joule heating for fixed electric field. 
 }
 \label{fig:Efix}
 \end{figure}

 So far, we have considered physical quantities for fixed $\Omega$.
 In Fig.~\ref{fig:Efix}, we show the Joule heatings as functions of $\Omega$ for several fixed $E$.
 When we take the slice of $E=\text{const.}$ in Fig.~\ref{fig:3dphasediagram}(a),
 the slice can cross several lobe-like regions in the insulator phase.
 In the lobe-like regions, the Joule heating is zero.
 Therefore, for small $E$ such as the curve for $E/m^2=0.01$ in the figure,
 the Joule heating becomes zero in several discrete regions.
 For  fields above the critical field, e.g., $E/m^2=0.06$,
 we still find a ``dip'' in the Joule heating around $\Omega/m\simeq 3.5$
 reflecting the lobe-shaped phase structure.

\section{Analytic derivation for features of the phase diagram in the high-frequency limit}
\label{sec:high_frequency}

When the frequency of the rotating electric fields is sufficiently high,
$\Omega \gg m$, the gauge field is well-described by linear
perturbations because $b(\rho) \sim \mathcal{O}(E/\Omega)$ (for example,
see Eq.~(\ref{bc_b})).
In this section, we study the high-frequency regime analytically.

We suppose that unperturbed background solutions are the vacuum state
without any electric field, 
that is, the trivial solutions $w(\rho) = m$ and $b(\rho) = 0$.
Perturbed solutions are given by 
$w(\rho) = 1 + \delta w$ and $b(\rho)\exp[i\chi(\rho)] = \delta b$.
For convenience, $\delta b$ will be treated as a complex function
including its phase and amplitude.
Note that we can set $m=1$ for the background solution
without loss of generality
because of the scaling symmetry (\ref{eq:scaling}) of the current systems.

The non-linear equations of motion (\ref{beq}) and (\ref{weq}) can be
linearized, so that the linearized equations of motion become 
for $\delta b$,
\begin{equation}
 \delta b'' + \frac{3}{\rho}\delta b'
  + \frac{\Omega^2}{(1+\rho^2)^2}\delta b = 0 ,
  \label{eq:eom_delta_b}
\end{equation}
and for $\delta w$,
\begin{equation}
 \delta w'' + \frac{3}{\rho}\delta w' = 
  \frac{2\Omega^2}{(1+\rho^2)^3}|\delta b|^2 .
  \label{eq:eom_delta_w}
\end{equation}
Since $\Omega$ is real, the complex conjugate $\delta b^*$ satisfies the
same equation of motion as (\ref{eq:eom_delta_b}). 

General solution of Eq.~(\ref{eq:eom_delta_b}) is given by 
\begin{equation}
 \delta b(\rho) = C_1 g_1(\rho) + C_2 g_2(\rho) ,
  \label{eq:general_solution1}
\end{equation}
where 
\begin{equation}
 \begin{aligned}
  g_1(\rho) =& (1+\rho^2)^\lambda F(\lambda,\lambda+1, 2; -\rho^2) ,\\
  g_2(\rho) =& \frac{1}{\rho^2}\left(1+\frac{1}{\rho^2}\right)^\lambda
  F(\lambda, \lambda+1, 2; -1/\rho^2) ,
 \end{aligned}
\end{equation}
and $\lambda \equiv (1+\sqrt{1+\Omega^2})/2$.
Each particular solution is a real function and has the following asymptotic behavior: 
\begin{equation}
 g_1(\rho) 
  =
  \left\{
   \begin{aligned}
    &\frac{1}{\Gamma(\lambda + 1)\Gamma(2-\lambda)}
    + \mathcal{O}(1/\rho^2)
    &
    (\rho \to \infty)
    \\
    &1 - \frac{\Omega^2}{8}\rho^2 + \mathcal{O}(\rho^4) \quad &
    (\rho \sim 0)
   \end{aligned}
  \right.
  ,
\end{equation}
and 
\begin{equation}
   g_2(\rho) =
   \left\{
   \begin{aligned}
    & \frac{1}{\rho^2} - \frac{\Omega^2}{8\rho^4} +
    \mathcal{O}(1/\rho^{6}) &
    (\rho \to \infty)
    \\
    & \frac{1}{\Gamma(\lambda + 1)\Gamma(2-\lambda)}
    \frac{1}{\rho^2} + \mathcal{O}(\rho^0\log\rho)
    &
    (\rho \sim 0)
   \end{aligned}
   \right.
   .
\end{equation}
Thus, in terms of asymptotic behavior near the AdS boundary
$\rho=\infty$, $g_1(\rho)$ and $g_2(\rho)$ describe so-called non-normalizable
modes and normalizable modes, respectively. 
At $\rho=0$ we find $g_1(\rho)$ is regular while $g_2(\rho)$ is singular.
This fact is closely related to whether the solution describes the
insulator or conductive phase (namely, the Minkowski or black hole embeddings).

For later convenience, using perturbative quantities we write down the conserved quantity $q$ defined by
Eq.~(\ref{qjule}), which corresponds to the Joule heating in the
boundary theory, as 
\begin{equation}
 \delta q = i \frac{\Omega}{2}
  \rho^3(\delta b \delta {b^*}' - \delta b^* \delta b') .
  \label{eq:delta_q}
\end{equation}
Since it is obviously proportional to the Wronskian of solutions of the linear
differential equation (\ref{eq:eom_delta_b}), $\delta q$ is 
conserved at this order as well as $q$ is conserved.
Substituting the general solution $\delta b(\rho)$ of
Eq.~(\ref{eq:general_solution2}), we explicitly obtain 
\begin{equation}
 \delta q = 
  \frac{8\cos\left(\frac{\pi}{2}\sqrt{1+\Omega^2}\right)}{\pi \Omega}
  \mathrm{Im}(C_2C_1^*) .
\end{equation}

General solution of Eq.~(\ref{eq:eom_delta_w}) is 
\begin{equation}
 \delta w(\rho) = - \frac{\Omega^2}{\rho^2}
  \int^\rho_0 ds \frac{s^3}{(s^2+1)^3}|\delta b(s)|^2
  - \Omega^2 \int^\infty_\rho ds
  \frac{s}{(s^2+1)^3}|\delta b(s)|^2 + C_3, 
  \label{eq:general_solution2}
\end{equation}
where $C_3$ is an integration constant equivalent to a boundary
condition $\delta w|_{\rho=\infty} = C_3$.

  \subsection{The insulator phase}
  
  As seen previously, the Minkowski embedding solution describing the
  insulator phase can reach to $\rho=0$.
  Since the solution should be regular at $\rho=0$, we choose $C_2 = 0$
  in Eq.~(\ref{eq:general_solution1}).
  Note that this leads to no Joule heating $\delta q =0$.
  The electric field $\delta E = \Omega \delta b|_{\rho=\infty}$ in the
  boundary theory is given by  
  \begin{equation}
   \delta E = \frac{C_1 \Omega}{\Gamma(\lambda+1)\Gamma(2-\lambda)}
    = - C_1 \frac{4\cos\left(\frac{\pi}{2}\sqrt{1+\Omega^2}\right)}
    {\pi \Omega} .
  \end{equation}
  It turns out that we have $\delta E = 0$ when $\sqrt{1+\Omega^2} =
  2n+3$ ($n=0,1,2,\ldots$), 
  of which frequencies correspond to the meson mass spectrum 
  $\Omega_n = 2\sqrt{(n+1)(n+2)}$ ($n=0,1,2,\ldots$).
  This means that at frequencies near the meson mass the insulator phase
  can be resonantly excited from the vacuum state with
  $E=0$ even by weak rotating electric fields.
  In the Minkowski embeddings, 
  $[\Omega b(\rho) - w^2(\rho)]|_{\rho=0} < 0$ should be satisfied, so
  that we have $\Omega\delta b|_{\rho=0} < 1$ for the perturbations.
  This implies that $C_1 < 1/\Omega$ and 
  $\delta E \sim \mathcal{O}(\Omega^{-2})$.

  The solution for $\delta w$ is given by Eq.~(\ref{eq:general_solution2}). 
  We require the boundary condition as $\delta w|_{\rho=0} = 0$, 
  which yields 
  \begin{equation}
   C_3 = \Omega^2 \int^\infty_0 ds \frac{s}{(s^2+1)^3}|\delta b(s)|^2
    = C_1^2 \Omega^2 \int^\infty_0 ds \frac{s}{(s^2+1)^3}g_1(s)^2 .
  \end{equation}
  From $w(\rho) = m + c/\rho^2 + \cdots$, the perturbations for the quark
  mass and the quark
  condensate can be
  read as 
  \begin{equation}
   \delta m = \epsilon^2 \int^\infty_0 ds \frac{s}{(s^2+1)^3}
    g_1(s)^2 ,
    \label{eq:delta_m_insulator}
  \end{equation}
  and 
  \begin{equation}
   \delta c = - \epsilon^2 
    \int^\infty_0 ds \frac{s^3}{(s^2+1)^3} g_1(s)^2 ,
    \label{eq:delta_c_insulator}
  \end{equation}
  where we have introduced a dimensionless parameter 
  $\epsilon \equiv \Omega C_1 < 1$.
  In addition, the electric field is 
  \begin{equation}
   \delta E 
    = - \epsilon \frac{4\cos\left(\frac{\pi}{2}\sqrt{1+\Omega^2}\right)}
    {\pi \Omega^2} .
    \label{eq:delta_E_insulator}
  \end{equation}

  If $\Omega=\Omega_n$ ($n=0,1,2,\ldots$), we can evaluate the
  integrals in $\delta m$ and $\delta c$ as 
  \begin{equation}
   \begin{aligned}
    \int^\infty_0 ds \frac{s}{(s^2+1)^3}g_1(s)^2 =&
    \frac{4n+3}{4(n+1)(n+2)(2n+3)} , \\
    \int^\infty_0 ds \frac{s^3}{(s^2+1)^3}g_1(s)^2 =&
    \frac{1}{4(n+1)(n+2)(2n+3)} .
   \end{aligned}
  \end{equation}
  Using the above results, 
  we have $\delta E \sim \Omega^{-2}$, $\delta m \sim \Omega^{-2}$, and
  $\delta c \sim \Omega^{-3}$ in the high-frequency regime.

  \subsection{The critical embeddings}
  
\begin{figure}
  \centering
  \includegraphics{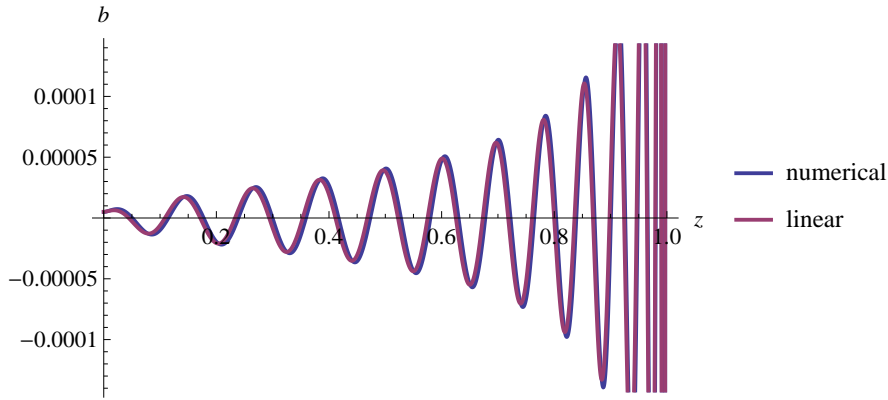}
  \caption{A critical solution for $b(\rho)$ obtained by numerically solving
  the non-linear full equation of motion and a
  linear perturbative solution of Eq.~(\ref{eq:delta_b_appro}) 
  for $\Omega = 50$.
  Note that we have used $z^2 = 1/(1+\rho^2)$ and $m=1$. }
  \label{fig:b_comparison}
 \end{figure}

\begin{figure}
  \centering
  \includegraphics{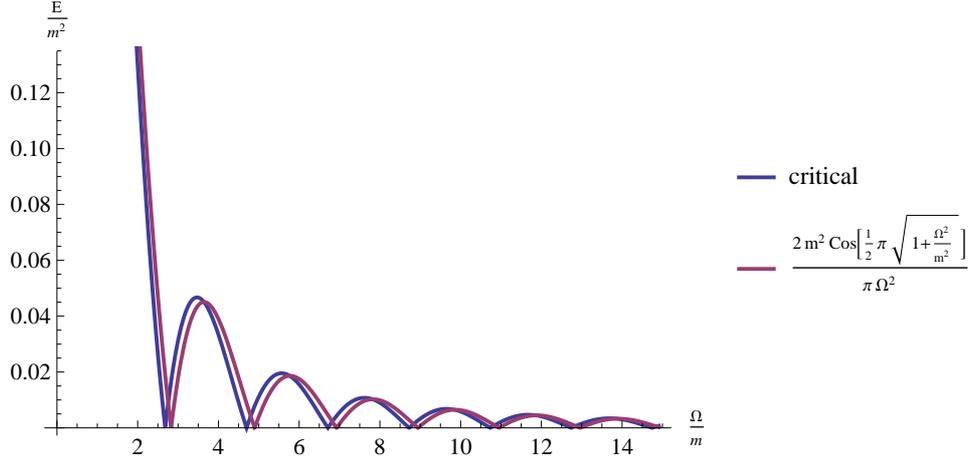}
  \caption{The electric field for the critical solutions.
  The red curve is an approximate analytic formula given by
  Eq.~(\ref{eq:approx_e}), and the blue curve is the critical value
  obtained numerically, which is identical to the phase boundary shown
  in Fig.~\ref{fig:3dphasediagram} (a).}
  \label{fig:E_comparison}
 \end{figure}

\begin{figure}
  \centering
  \includegraphics{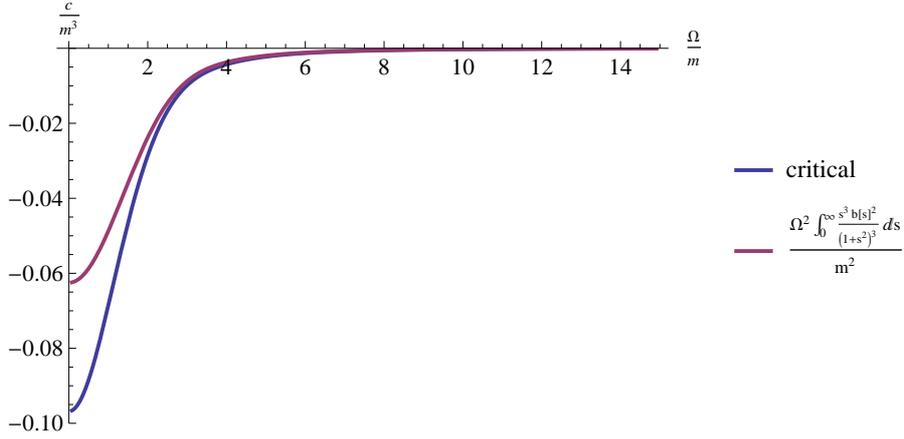}
  \caption{The quark condensate for the critical solutions. The red
  curve is an approximate formula given by Eq.~(\ref{eq:approx_c}), and
  the blue curve is the critical value obtained numerically.}
  \label{fig:c_comparison}
 \end{figure}

  Since the critical solutions satisfy 
  $[\Omega b(\rho) - w^2(\rho)]|_{\rho=0} =0$, 
  the limit $\epsilon \to 1$ seems to indicate the perturbed solutions
  may become critical. 
  Actually, it is not true because of non-linearity.
  As seen in Eqs.~(\ref{baxis_c}) and (\ref{waxis_c}), the critical
  solutions behave linearly with respect to $\rho$ near $\rho=0$.
  On the other hand, the linearized solutions are quadratic except for
  divergent one.
  This implies that for the critical solutions non-linearity has a
  significant role around $\rho = 0$, so that they cannot be described
  by the linear perturbations but higher-order corrections should be necessary.
  In a strict sense, the linear solutions of the conductive phase described
  previously are valid for $\epsilon \ll 1$.
  However, it turns out that for $\rho \gg 1$ the solutions of linear
  perturbations can describe the true non-linear solution very well.
  Indeed, if we assume 
  \begin{equation}
   \delta b(\rho) = \frac{1}{2\Omega} g_1(\rho) ,
    \label{eq:delta_b_appro}
  \end{equation}
  which is identical to Eq.~(\ref{eq:general_solution1}) with
  $C_1=1/(2\Omega)$ (namely, $\epsilon=1/2$) and $C_2=0$, we explicitly
  observe that the perturbative solution is quite similar to the non-linear
  solution for almost every $\rho$ as shown in Fig.~\ref{fig:b_comparison}.
 Thus, if we choose $\epsilon=1/2$ in Eqs.~(\ref{eq:delta_c_insulator})
 and (\ref{eq:delta_E_insulator}), we can approximately evaluate the
 critical electric field and the critical quark condensate as 
 \begin{equation}
  |E_\mathrm{approx}| = \frac{2}{\pi\Omega^2}
   \left|\cos\frac{\pi}{2}\sqrt{1+\Omega^2}\right| \quad
   (\Omega \gg 1) ,
   \label{eq:approx_e}
 \end{equation}
 and 
 \begin{equation}
  c_\mathrm{approx} = - \frac{1}{4}\int^\infty_0 ds
   \frac{s^3}{(s^2+1)^3} g_1(s)^2 \quad (\Omega \gg 1).
   \label{eq:approx_c}
 \end{equation}
 Comparing the approximate formulae (\ref{eq:approx_e}) and
 (\ref{eq:approx_c}) with the numerical solutions, we find that in
 the high-frequency regime the numerical solutions can be fitted by the
 approximate formulae as shown in Fig.~\ref{fig:E_comparison} and
 Fig.~\ref{fig:c_comparison}.
 Several discrepancies imply that non-linearity plays a
 significant role.
 As a result, we can confirm that the critical electric field has a
 dependence of $\Omega^{-2}$ as the frequency becomes large, and it
 becomes periodically infinitesimal near frequencies of the meson spectrum.

  \subsection{The conductive phase}

  The conductive phase after the dielectric breakdown is described by
  the black hole embedding solution with the effective horizon on the
  D$7$-brane.
  In this phase the gauge field on the D$7$-brane becomes a propagating
  wave with a net energy flux from the AdS boundary $\rho=\infty$ 
  to the effective horizon $\rho=\rho_c$ such that 
  $[\Omega b(\rho) - w^2(\rho)]|_{\rho=\rho_c}=\rho_c^2$.
  This means the perturbative solution $\delta b(\rho)$ should represent
  a purely-ingoing wave without any reflection at $\rho=\rho_c$.

  In order to clarify the boundary condition, 
  we redefine a coordinate and a variable as  
  $\psi(y) \equiv \frac{\rho^{3/2}}{\sqrt{1+\rho^2}} \delta b$ 
  and $\rho=\tan y$, so that Eq.~(\ref{eq:eom_delta_b}) becomes
  \begin{equation}
   \left[\frac{d^2}{dy^2} + \Omega^2 - U(y)\right]\psi(y) = 0 ,
    \quad U(y) \equiv \frac{5+\cos(4y)}{2\sin^2(2y)} .
  \end{equation}
  By introducing $C_1 = \alpha$ and $C_2 =\beta + i\gamma$
  ($\alpha,\beta,\gamma \in \mathbb{R}$)%
  \footnote{Since $\delta b(\rho)$ admits arbitrary constant phase
  shift, we can assume $C_1$ is real without loss of generality.}
  , its general solution is given
  by 
  \begin{equation}
   \psi(y) = \alpha \hat{g}_1(y) + (\beta + i\gamma) \hat{g}_2(y) ,
  \end{equation}
  where each particular solution $\hat{g}_i(y)$ ($i=1,2$) is redefined
  by $\hat{g}_i(y) \equiv \rho^{3/2}(1+\rho^2)^{-1/2} g_i(\rho)$.
  Since the locus of the effective horizon is given by 
  $\Omega|\delta b(\rho_c)| = 1+\rho_c^2$ at the linear order, 
  we have a condition 
  \begin{equation}
   |\psi(y_0)|^2 = \frac{\sin^3 y_0}{\Omega^2 \cos^5 y_0} ,
    \label{eq:horizon_condition}
  \end{equation}
  where $0< y_0 \equiv \tan^{-1}\rho_c < \pi/2$.
  Note that in order to satisfy $|\delta b| \ll 1$ we assume 
  $\cos^2 y_0 \gg 1/\Omega$ at least. 
  In addition, purely-ingoing boundary condition%
  \footnote{It implies that the complex gauge field should behave as 
  $a(t,\rho(y)) \sim e^{i\Omega(t+y-y_0)}$ around $y=y_0$.}
  at $y=y_0$ is given by  
  \begin{equation}
   \left.\frac{d\psi}{dy}\right|_{y=y_0} = i\Omega\psi(y_0).
    \label{eq:transmitting_condition}
  \end{equation}
  Thus, the above conditions (\ref{eq:horizon_condition}) and
  (\ref{eq:transmitting_condition}) can determine the coefficients
  $\alpha$, $\beta$, and $\gamma$ for a given $y_0$.
  Note that since the Joule heating (\ref{eq:delta_q}) is rewritten as 
  \begin{equation}
   \delta q = i \frac{\Omega}{2}
    \left(\psi\frac{d\psi^*}{dy} - \psi^*\frac{d\psi}{dy}\right) ,
  \end{equation}
  those conditions yield 
  $\delta q=\Omega^2|\psi(y_0)|^2 = \rho_c^3(1+\rho_c^2)$ 
  at $y=y_0$.
  This $\delta q$ is equivalent to Eq.~(\ref{q_horizon}) in fully
  non-linear cases.

  Now, we have 
  \begin{equation}
   \alpha = 
    \frac{\Omega^2 (\hat{g}_2)^2 + \left(\frac{d}{dy}\hat{g}_2\right)^2}
    {\Omega W(\hat{g}_1,\hat{g}_2)}
    \gamma,
    \quad
    \beta = - \frac{\Omega^2 \hat{g}_1\hat{g}_2
    + \left(\frac{d}{dy}\hat{g}_1\right)\left(\frac{d}{dy}\hat{g}_2\right)}
    {\Omega W(\hat{g}_1,\hat{g}_2)}\gamma ,
  \end{equation}
  where 
  $W(\hat{g}_1,\hat{g}_2) \equiv \hat{g}_1 \frac{d}{dy}\hat{g}_2 - \hat{g}_2
  \frac{d}{dy}\hat{g}_1 = \frac{8}{\pi\Omega^2}\cos\left(\frac{\pi}{2}\sqrt{1+\Omega^2}\right)$ is the Wronskian of $\hat{g}_1(y)$ and $\hat{g}_2(y)$,
  and also we have 
  \begin{equation}
   \gamma^2 = \frac{\sin^3 y_0}
    {\cos^5 y_0 \left[\Omega^2 (\hat{g}_2)^2
	       + \left(\frac{d}{dy}\hat{g}_2\right)^2\right]} .
  \end{equation}

  The Joule heating is 
  \begin{equation}
   \delta q = 
    \frac{8}{\pi\Omega}\cos\left(\frac{\pi}{2}\sqrt{1+\Omega^2}\right)
    \alpha\gamma
    = \frac{\sin^3 y_0}{\cos^5 y_0} ,
  \end{equation}
  and the electric field is 
  \begin{equation}
   \delta E = - \frac{4}{\pi\Omega}
    \cos\left(\frac{\pi}{2}\sqrt{1+\Omega^2}\right)
    \alpha
    = - \frac{1}{2}\sqrt{\frac{\sin^3 y_0}{\cos^5 y_0}
    \left[\Omega^2 (\hat{g}_2)^2 + \left(\frac{d\hat{g}_2}{dy}\right)^2\right]} .
  \end{equation}

\section{Toy model calculation (2d QED)}
\label{sec:toy}

In this section, we perform a complementary field theoretical calculation 
to understand the universal nature of the dielectric breakdown of
charge confined systems excited by oscillating fields. 
The system we start with is the (1+1)-dimensional quantum electrodynamics, 
also known as the massive Schwinger model (see Ref.~\cite{Coleman:1976} and references therein for details). 
We apply  an AC-electric field to this system. 
The  defining Lagrangian density is given by 
\begin{equation}
\mathcal{L}=-\frac{1}{4}F_{\mu\nu}F^{\mu\nu}+\bar{\psi}(i\Slash{\pa}-e\Slash{A}-m)\psi
\ .
  \label{mSM}
\end{equation}
The model is define on a (1+1)-dimensional spacetime ($\mu=0,1$) and the field strength is given by $F_{\mu\nu}=\pa_\mu A_\nu-\pa_\nu A_\mu$. The charge $e$ and mass $m$ are positive numbers and we set $e=1$. 
The system is in the confined phase in the ground state; There are no free quarks
but only mesons, i.e. bound state of quarks and antiquarks, exist in the excitation spectrum. 
The merit of studying this models is that we can study its strong coupling limit through  bosonization. 
 It is mapped to a model  of a non-compact boson field $\phi(t,x)$ with a Hamiltonian density
\begin{equation}
\mathcal{H}=N_m\left[\frac{1}{2}
\Pi^2+\frac{1}{2}(\pa_1\phi)^2-cm^2\cos 2\pi^{1/2}\phi+\frac{1}{2\pi}\left(\phi+\frac{1}{2\pi^{1/2}}\theta\right)^2
\right]
\ ,
  \label{mSMb}
\end{equation}
where $N_m$ denotes normal ordering, $\Pi$ the conjugate field of $\phi$, $c$ a numerical constant. 
The parameter $\theta$ is proportional to the external electric field, 
which we set to be oscillating with frequency $\Omega$
\begin{equation}
\theta=2\pi E\cos\Omega t
\ .
  \label{theta}
\end{equation}
The bosonic model becomes reliable  in the strong coupling $e(=1)\gg m$ limit. 
The confining nature of the model can be seen by noticing that the 
fermionic excitation with a charge is given by a kink connecting the minima
of the cosine potential, e.g., from $\phi=0$ to $\phi=\pm \pi^{1/2}$. Due
to the quadratic potential, such excitation will be energetically suppressed since $\phi=\pm \pi^{1/2}$ will only be a local minima. 
If we have a kink and anti-kink pair separated by a length $l$, the potential 
energy linearly increases as $V(l)=l/2$ leading to charge confinement.

\begin{figure*}[tb]
\centering 
\includegraphics[scale=0.65]{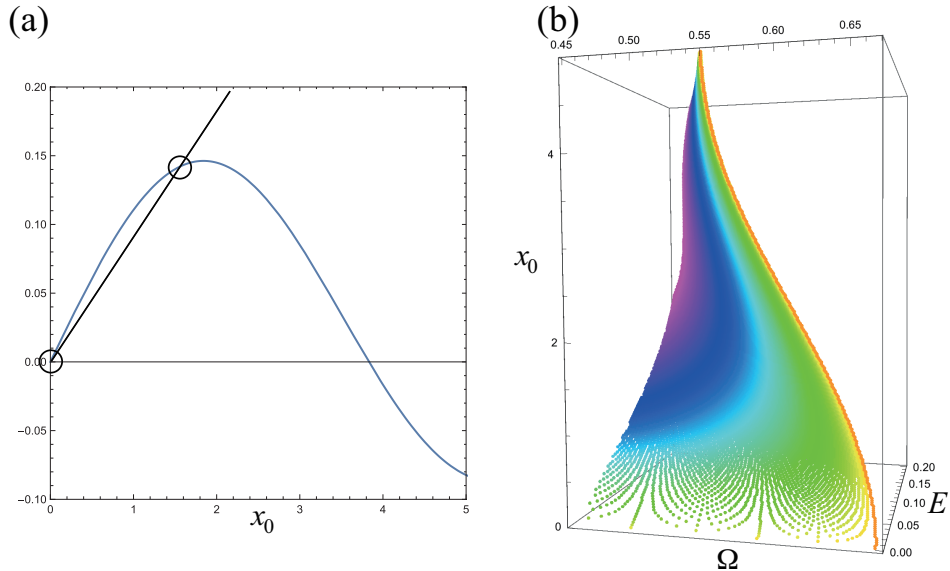}
\caption{Analysis of the toy model: (a) The left and right hand sides of the equation (\ref{eq:toymodel}), 
giving multiple solutions denoted by circles for zero field $E=0$. Note that the numerical constant $c$ is set to unity. 
(b) The surface of solutions. The orange line corresponds to solutions at $E=0$. 
}
\label{fig:toysolve}
\end{figure*}

Below, we neglect spatial and quantum fluctuations, and simply  treat it as a problem of classical dynamics governing
the zero mode $\bra \phi(x,t)\ket=\phi (t)$. 
The equation of motion becomes
\begin{equation}
\ddot{\phi}+\pi^{-1}\phi+cm^22\pi^{1/2}\sin 2\pi^{1/2}\phi=-\pi^{-1/2}E\cos\Omega t
\end{equation}
This is nothing but a model of a periodically driven anharmonic oscillator. 
Assuming a simple oscillatory solution  $\phi=\phi_0\cos\Omega t$ and using an ansatz $\sin(a\cos x)\sim 2 J_1(a)\cos x+\ldots$,
we obtain a relation 
\begin{equation}
cm^28\pi J_1( x_0)=(\Omega^2-\pi^{-1})x_0+2E
\label{eq:toymodel}
\end{equation}
that determines the amplitude of the oscillation ($J_1$ is the 1st-Bessel function).  
Note that we defined $x_0=2\pi^{1/2}\phi_0$. 
This equation can have multiple solutions for a given parameter set $(E,\Omega)$  (Fig.~\ref{fig:toysolve}(a)).
In this model, $\phi_0$ represents the condensation of quark-antiquark pairs (=meson). 
Thus, in the toy model,
(i) We find multiple solution for a given parameter $(\Omega,E)$, and (ii) We find a state with finite $\phi_0$ existing even at $E=0$ (meson condensate), and 
(iii) The resonant frequency depends on the amplitude of the oscillation. 
The current also oscillates with the same frequency as the condensation
and the behavior of the orange line shown in Fig.~\ref{fig:toysolve}(b)
is similar to what we obtained in the D3/D7 system plotted in Fig.~\ref{fig:3dphasediagram}(a),~(b). 
Thus, this simple toy model partially captures aspects of the meson condensate;
In fact, the resonant frequency being decreased as the amplitude increases can be 
attributed to the anharmonicity of the potential and is well known in the study of 
Duffin oscillators, i.e., a oscillator with 2nd and 4th term potential. 
Important properties that are {\em not} captured by this model is the chiral nature. In the 
Floquet condensate of vector mesons in rotating electric fields, the rotating current breaks time reversal symmetry. 
In addition, although the current is time dependent, the scalar meson condensation is static. 
These features are unique to the Floquet condensate of vector mesons and deserve further field theoretical understanding.

\begin{acknowledgments}
 We acknowledge discussion with Haruki Watanabe and Markus Heyl.
We would especially like to thank Koji Hashimoto for collaboration at an earlier stage.
 This work was partially supported by JSPS KAKENHI Grant Numbers JP16K17704 (S. K.)
 and by ImPACT Program of Council for Science, Technology and Innovation, Cabinet Office, Government of Japan (Grant No. 2015-PM12-05-01) (T. O.).
The work was supported by JSPS KAKENHI Grant Number 15K17658 (K. M.). 
\end{acknowledgments}

\appendix

\section{Boundary conditions at the effective horizon}
\label{app:BC}

In the equations of motion~(\ref{beq}), (\ref{chieq}) and (\ref{weq}),
the denominators of the right-hand side become zero at the effective horizon $\rho=\rho_c$.
For the regularity of the second derivatives of $b$, $\chi$ and $w$,
the numerators in those equations should also be zero at $\rho=\rho_c$.
Using $w_0^2 + \rho_c^2 = \Omega b_0$,
we obtain regularity conditions as
  \begin{align}
   & 
   - \rho_c(2\rho_c b' + \Omega)(1+w'^2+b'^2)
   + \Omega b_0^2 (\rho_c - 3b_0 b')\chi'^2 = 0, \label{hor1} 
   \\
   & 
    - 
   \left[
   2\rho_c^2(-1+w'^2+b'^2)
   - 
   2\rho_c \left(2 w_0 w' - \Omega b'\right)
   \right]\chi'
   -3\Omega b_0^3 \chi'^3
   = 0, \label{hor2} 
   \\
   &
   - 2\rho_c \left(\rho_c w' - w_0\right)(1+w'^2+b'^2)
   - 3 \Omega b_0^3 w' \chi'^2
   = 0. \label{hor3} 
  \end{align}
  From Eq.~(\ref{hor1}),
  we can express $\chi'$ in terms of $b'$ and $w'$ as
  \begin{equation}
   \chi'^2 = - \frac{\rho_c(2\rho_c b' + \Omega)(1+w'^2+b'^2)}
    {\Omega b_0^2 (3b_0 b' - \rho_c)} .
    \label{chiprime}
  \end{equation}
  Eliminating $\chi'$ from Eq.~(\ref{hor3}) using the above equation,
  we obtain
  \begin{equation}
   b' = - \frac{(2\rho_c^2 + 3\Omega b_0)w' - 2\rho_c w_0}
    {6 w_0 b_0} . 
    \label{bprime}
  \end{equation}  
  Eliminating $b'$ and $\chi'$ from Eq.~(\ref{hor2}) using
  Eqs.~(\ref{chiprime}) and (\ref{bprime}), we have an equation for $w'$
  as
  \begin{equation}
   \left\{1+\frac{\Omega^2(w_0^2+\rho_c^2)(9w_0^2+5\rho_c^2)}
    {4w_0^2[9(w_0^2+\rho_c^2)^2 + \Omega^2\rho_c^2]}\right\}w'^2
   +\frac{2\rho_c}{w_0}w' - 1=0\ .
   \label{wprime}
  \end{equation}
  As a result, we obtain a solution of Eqs.~(\ref{hor1})-(\ref{hor3}) as
\begin{align}
 w' =& \frac{2w_0}{D}
 \left[- 18\rho_c (w_0^2 + \rho_c^2)^2 - 2\Omega^2\rho_c^3
 + 3(w_0^2 + \rho_c^2)s
 \right] \ ,
 \\
 b' =& \frac{\Omega}{D}
 \left[
 30\rho_c(w_0^2+\rho_c^2)^2 + \Omega^2\rho_c(5\rho_c^2+3w_0^2)
 - (5\rho_c^2+3w_0^2)s
 \right] \ ,
 \\
 \chi'^2 =& \frac{2\Omega^2\rho_c
 \left[4(w_0^2 + \rho_c^2) + \Omega^2\right]}
 {D^2 (w_0^2 + \rho_c^2)^2}
 \left\{
 - \rho_c \left[216w_0^2(w_0^2+\rho_c^2)^3
 \right.
 \right.
 \nonumber\\
 &
 \left.
 + 6\Omega^2(w_0^2+\rho_c^2)(15w_0^4 + 44w_0^2\rho_c^2 + 25\rho_c^4)
 + \Omega^4 (5\rho_c^2+3w_0^2)^2
 \right]
 \\
 &
 \left.
 + \left[36w_0^2(w_0^2+\rho_c^2)^2 + \Omega^2(5\rho_c^2+3w_0^2)^2\right]s
 \right\}
 \nonumber
\end{align}
where
\begin{equation}
 \begin{aligned}
  s \equiv& \sqrt{\left[4(w_0^2 + \rho_c^2) + \Omega^2\right]
  \left[9(w_0^2 + \rho_c^2)^2 + \Omega^2 \rho_c^2\right]
  } \ ,
  \\
  D \equiv& 36 w_0^2 (w_0^2+\rho_c^2)^2
  + \Omega^2 (5\rho_c^4 + 18\rho_c^2w_0^2 + 9w_0^4) .
 \end{aligned}
\end{equation}

Note that, although (\ref{wprime}) has two roots, we have taken the root
for $s>0$ in the above solutions.
In the static limit $\Omega\to 0$ they are reduced to
\begin{equation}
 \begin{aligned}
  w' \to \frac{- \rho_c + \sqrt{w_0^2 + \rho_c^2}}{w_0}, \quad
  b' \to 0 , \quad
  \chi'^2/\Omega^2 \to \frac{4\rho_c}{3w_0^2(w_0^2+\rho_c^2)^2}
  \left(-\rho_c + \sqrt{w_0^2 + \rho_c^2}\right) ,
 \end{aligned}
\end{equation}
which are usual boundary conditions in the case of static electric fields.
If we choose the solutions for $s<0$, the boundary conditions are
singular at the massless limit $w_0\to 0$.


\begin{thebibliography}{99}

\bibitem{Schwinger:1951} 
  J.~S.~Schwinger,
  ``On gauge invariance and vacuum polarization,''
  Phys.\ Rev.\  {\bf 82}, 664 (1951).

\bibitem{Heisenberg1936} 
W. Heisenberg, and H. Euler, ``Folgerungen aus der Diracschen Theorie des Positrons'', Z. Phys. {\bf 1} 714--732 (1936), 

\bibitem{Weisskopf}  
V. Weisskopf, 
``The electrodynamics of the vacuum based on the quantum theory of the electron,'' 
In *Miller, A.I.: Early quantum electrodynamics* 206-226

\bibitem{Maldacena:1997re} 
  J.~M.~Maldacena,
  ``The Large N limit of superconformal field theories and supergravity,''
  Adv.\ Theor.\ Math.\ Phys.\  {\bf 2}, 231 (1998)
  [hep-th/9711200].

\bibitem{Gubser:1998bc} 
  S.~S.~Gubser, I.~R.~Klebanov and A.~M.~Polyakov,
  ``Gauge theory correlators from noncritical string theory,''
  Phys.\ Lett.\ B {\bf 428}, 105 (1998)
  [hep-th/9802109].
 
\bibitem{Witten:1998qj} 
  E.~Witten,
  ``Anti-de Sitter space and holography,''
  Adv.\ Theor.\ Math.\ Phys.\  {\bf 2}, 253 (1998)
  [hep-th/9802150].


\bibitem{Hashimoto:2016ize} 
  K.~Hashimoto, S.~Kinoshita, K.~Murata and T.~Oka,
  ``Holographic Floquet states: (I) A strongly coupled Weyl semimetal,''
  arXiv:1611.03702 [hep-th].


\bibitem{Karch}
A.~Karch and E.~Katz, 
``Adding flavor to AdS / CFT,'' 
	JHEP {\bf 0206} 043 (2002), [arXiv:hep-th/0205236].

\bibitem{Kruczenski:2003be} 
  M.~Kruczenski, D.~Mateos, R.~C.~Myers and D.~J.~Winters,
  ``Meson spectroscopy in AdS / CFT with flavor,''
  JHEP {\bf 0307}, 049 (2003)
  [hep-th/0304032].

\bibitem{Karch:2007pd} 
  A.~Karch and A.~O'Bannon,
  ``Metallic AdS/CFT,''
  JHEP {\bf 0709}, 024 (2007)
  [arXiv:0705.3870 [hep-th]].

\bibitem{Albash:2007bq} 
  T.~Albash, V.~G.~Filev, C.~V.~Johnson and A.~Kundu,
  ``Quarks in an external electric field in finite temperature large N gauge theory,''
  JHEP {\bf 0808}, 092 (2008)
  [arXiv:0709.1554 [hep-th]].

\bibitem{Erdmenger:2007bn} 
  J.~Erdmenger, R.~Meyer and J.~P.~Shock,
  ``AdS/CFT with flavour in electric and magnetic Kalb-Ramond fields,''
  JHEP {\bf 0712}, 091 (2007)
  [arXiv:0709.1551 [hep-th]].

\bibitem{Nakamura:2010zd} 
  S.~Nakamura,
  ``Negative Differential Resistivity from Holography,''
  Prog.\ Theor.\ Phys.\  {\bf 124}, 1105 (2010)
  doi:10.1143/PTP.124.1105
  [arXiv:1006.4105 [hep-th]].

\bibitem{Nakamura:2012ae} 
  S.~Nakamura,
  ``Nonequilibrium Phase Transitions and Nonequilibrium Critical Point from AdS/CFT,''
  Phys.\ Rev.\ Lett.\  {\bf 109}, 120602 (2012)
  [arXiv:1204.1971 [hep-th]].

\bibitem{Nakamura:2013yqa} 
  S.~Nakamura and H.~Ooguri,
  ``Out of Equilibrium Temperature from Holography,''
  Phys.\ Rev.\ D {\bf 88}, no. 12, 126003 (2013)
  [arXiv:1309.4089 [hep-th]].

\bibitem{Hashimoto:2014yza} 
  K.~Hashimoto, S.~Kinoshita, K.~Murata and T.~Oka,
  ``Electric Field Quench in AdS/CFT,''
  JHEP {\bf 1409}, 126 (2014)
  [arXiv:1407.0798 [hep-th]].

\bibitem{Hoshino:2014nfa} 
  H.~Hoshino and S.~Nakamura,
  ``Effective temperature of nonequilibrium dense matter in holography,''
  Phys.\ Rev.\ D {\bf 91}, no. 2, 026009 (2015)
  [arXiv:1412.1319 [hep-th]].







\bibitem{Dunne2004}
Gerald V. Dunne, hep-th/0406216. 

\bibitem{Gelis:review}
F.~Gelis, N.~Tanji, ``Schwinger mechanism revisited,''
Prog. Particle and Nuclear Phys. {\bf 87}, 1 (2016)



\bibitem{Gorsky:2011}
A. S. Gorsky, K. A. Saraikin and K. G. Selivanov, 
``Schwinger type processes via branes and their gravity duals,''
Nucl. Phys. {\bf B 628}, 270 (2002)  [hep-th/0110178].

\bibitem{SemenoffZarembo}
G.W. Semenoff, K. Zarembo, 
Phys. Rev. Lett. {\bf 107}, 171601 (2011), arXiv:1109.2920 [hep-th]. 

\bibitem{AmbjornMakeenko}
 J. Ambjorn, Y. Makeenko, 
 Phys. Rev. {\bf D 85} 061901 (2012).
arXiv:1112.5606 [hep-th]. 
 
 \bibitem{Bolognesi}
 S. Bolognesi, F. Kiefer, E. Rabinovici, 
 J. High Energy Phys. {\bf 01}, 174 (2013)
arXiv:1210.4170 [hep-th].
 
 \bibitem{Sato:2013}
 Y. Sato, K. Yoshida, J. High Energy Phys. {\bf 04}, 111 (2013)
arXiv:1303.0112 [hep-th].
 
 \bibitem{Sato:20132}
 Y. Sato, K. Yoshida, 
 J. High Energy Phys. {\bf 08}, 002 (2013) 
arXiv:1304.7917 [hep-th].

\bibitem{Dietrich2014}
D.D. Dietrich, Phys. Rev. {\bf D 90},  045024 (2014)
arXiv:1405.0487 [hep-ph].

\bibitem{Kawai:2015}
D. Kawai, Y. Sato, K. Yoshida, Internat. J. Modern Phys. {\bf A 30} 1530026 (2015)
arXiv:1504.00459 [hep-th]

\bibitem{Hashimoto:2013}
K. Hashimoto, T. Oka, J. High Energy Phys. {\bf 10}, 116  (2013)
arXiv:1307.7423 [hep-th].

\bibitem{Hashimoto:2015}
K. Hashimoto, T. Oka, A. Sonoda, J. High Energy Phys. {\bf 06} 001 (2015)
arXiv:1412.4254 [hep-th].

\bibitem{Sonner:2013}
J. Sonner, Phys. Rev. Lett. {\bf 111}, 211603 (2013)
arXiv:1307.6850 [hep-th]. 


\bibitem{Floquetcondensate}
C. Heinisch, M. Holthaus,
``Adiabatic preparation of Floquet condensates'',
J. Mod. Opt. {\bf 63}, 1768 (2016).

\bibitem{Murakami17}
Y. Murakami, N. Tsuji, M. Eckstein, P. Werner,
``Nonequilibrium steady states and transient dynamics of conventional superconductors under phonon driving'',
Phys. Rev. B {\bf 96}, 045125 (2017).

\bibitem{Knap16}
M. Knap, M. Babadi, G. Refael, I. Martin, E. Demler,
``Dynamical Cooper pairing in non-equilibrium electron-phonon systems''
 Phys. Rev. B {\bf 94}, 214504 (2016)

\bibitem{Babadi17}
M. Babadi, M. Knap, I. Martin, G. Refael, E. Demler,
``The theory of parametrically amplified electron-phonon superconductivity'',
Phys. Rev. B {\bf 96}, 014512 (2017)

\bibitem{Wilczek:2012} 
  F.~Wilczek,
  ``Quantum Time Crystals,''
  Phys.\ Rev.\ Lett.\  {\bf 109}, 118901 (2012)
  
\bibitem{Berges:2004} 
J. Berges, Sz. Borsanyi, and C.Wetterich,
  ``Prethermalization,''
  Phys.\ Rev.\ Lett.\  {\bf 93}, 142002 (2004)



\bibitem{Khemani:2016}
V. Khemani, A. Lazarides, R. Moessner, S. L. Sondhi, ``On the phase structure of driven quantum systems'', Phys. Rev. Lett. {\bf 116}, 250401 (2016)

\bibitem{Else:2016}
Dominic V. Else, Bela Bauer, Chetan Nayak, ``Floquet Time Crystals'', arXiv:1603.08001, (2016)



 \bibitem{ZhangTC:2016}
J. Zhang, P. W. Hess, A. Kyprianidis, P. Becker, A. Lee, J. Smith, G. Pagano, I.-D. Potirniche, A. C. Potter, A. Vish- wanath, N. Y. Yao, C. Monroe, ``Observation of a Discrete Time Crystal'', arXiv:1609.08684, (2016)

\bibitem{ChoiTC:2016}
Soonwon Choi, Joonhee Choi, Renate Landig, Georg Kuc- sko, Hengyun Zhou, Junichi Isoya, Fedor Jelezko, Shi- nobu Onoda, Hitoshi Sumiya, Vedika Khemani, Curt von Keyserlingk, Norman Y. Yao, Eugene Demler, Mikhail D. Lukin, ``Observation of discrete time-crystalline order in a disordered dipolar many-body system'', arxiv:1610.08057, (2016)

\bibitem{Yoshioka:2011}
K.~Yoshioka, E.~ Chae, and M.~Kuwata-Gonokami,
``Transition to a Bose-Einstein condensate and relaxation explosion of excitons at sub-Kelvin temperatures,''
Nat. Com. {\bf 2}, 328 (2011).

\bibitem{Bismuthreview}
Y.~Fuseya, M.~Ogata2, and H.~Fukuyama, 
``Transport Properties and Diamagnetism of Dirac Electrons in Bismuth'', 
Journal of the Physical Society of Japan 84, 012001 (2015).

\bibitem{Murakami_exciton18}
Yuta Murakami, Denis Golez, Martin Eckstein, and Philipp Werner, arXiv:1707.07706. 

\bibitem{Tanay18}
Tanay Nag, Robert-Jan Slager, Takuya Higuchi, and Takashi Oka, {\em in preparation}.


\bibitem{TI1}
M. Z. Hasan, C. L. Kane, 
Rev. Mod. Phys. {\bf 82}, 3045 (2010).

\bibitem{TI2}
X. L. Qi, S. C. Zhang, Rev. Mod. Phys. {\bf 83}, 1057 (2011).


\bibitem{Gedik1}
Y. H. Wang, H. Steinberg, P. Jarillo-Herrero, N. Gedik, 
Science {\bf 342}, 453 (2013).


\bibitem{Suzuki:2012} 
T.~Suzuki and R.~Shimano,
``Exciton Mott Transition in Si Revealed by Terahertz Spectroscopy,''
Phys. Rev. Lett. {\bf 109}, 046402 (2012).

\bibitem{Mott:1968}
N. F. Mott, Rev. Mod. Phys. {\bf 40}, 677 (1968). 

\bibitem{Zimmermann:1988}
R. Zimmermann, Many-Particle Theory of Highly Excited Semiconductors (Teubner, Leipzig, 1988).

\bibitem{He:mesomMott}
Lianyi He,
``Nambu-Jona-Lasinio model description of weakly interacting Bose condensate and BEC-BCS crossover in dense QCD-like theories,''
Phys. Rev.{\bf D 82}, 096003 (2010).

\bibitem{Kharzeev:2007jp} 
  D.~E.~Kharzeev, L.~D.~McLerran and H.~J.~Warringa,
  ``The Effects of topological charge change in heavy ion collisions: 'Event by event P and CP violation',''
  Nucl.\ Phys.\ A {\bf 803}, 227 (2008)
  [arXiv:0711.0950 [hep-ph]].
  
\bibitem{Skokov:2009qp} 
  V.~Skokov, A.~Y.~.Illarionov and V.~Toneev,
  ``Estimate of the magnetic field strength in heavy-ion collisions,''
  Int.\ J.\ Mod.\ Phys.\ A {\bf 24}, 5925 (2009)
  [arXiv:0907.1396 [nucl-th]].



\bibitem{Voronyuk:2011jd} 
  V.~Voronyuk, V.~D.~Toneev, W.~Cassing, E.~L.~Bratkovskaya, V.~P.~Konchakovski and S.~A.~Voloshin,
  ``(Electro-)Magnetic field evolution in relativistic heavy-ion collisions,''
  Phys.\ Rev.\ C {\bf 83}, 054911 (2011)
  [arXiv:1103.4239 [nucl-th]].
  
\bibitem{Bzdak:2011yy} 
  A.~Bzdak and V.~Skokov,
  ``Event-by-event fluctuations of magnetic and electric fields in heavy ion collisions,''
  Phys.\ Lett.\ B {\bf 710}, 171 (2012)
  [arXiv:1111.1949 [hep-ph]].


\bibitem{Deng:2012pc} 
  W.~-T.~Deng and X.~-G.~Huang,
  ``Event-by-event generation of electromagnetic fields in heavy-ion collisions,''
  Phys.\ Rev.\ C {\bf 85}, 044907 (2012)
  [arXiv:1201.5108 [nucl-th]].

\bibitem{Frolov:2006tc} 
  V.~P.~Frolov,
  ``Merger Transitions in Brane-Black-Hole Systems: Criticality, Scaling, and Self-Similarity,''
  Phys.\ Rev.\ D {\bf 74}, 044006 (2006)
  [gr-qc/0604114].


\bibitem{Mateos:2006nu} 
  D.~Mateos, R.~C.~Myers and R.~M.~Thomson,
  ``Holographic phase transitions with fundamental matter,''
  Phys.\ Rev.\ Lett.\  {\bf 97}, 091601 (2006)
  [hep-th/0605046].

\bibitem{Mateos:2007vn} 
  D.~Mateos, R.~C.~Myers and R.~M.~Thomson,
  ``Thermodynamics of the brane,''
  JHEP {\bf 0705}, 067 (2007)
  [hep-th/0701132].

\bibitem{Seiberg:1999vs} 
  N.~Seiberg and E.~Witten,
  ``String theory and noncommutative geometry,''
  JHEP {\bf 9909}, 032 (1999)
  [hep-th/9908142].

\bibitem{Gibbons:2000xe} 
  G.~W.~Gibbons and C.~A.~R.~Herdeiro,
  ``Born-Infeld theory and stringy causality,''
  Phys.\ Rev.\ D {\bf 63}, 064006 (2001)
  [hep-th/0008052].

\bibitem{Gibbons:2001ck} 
  G.~W.~Gibbons,
  ``Pulse propagation in Born-Infeld theory: The World volume equivalence principle and the Hagedorn - like equation of state of the Chaplygin gas,''
  Grav.\ Cosmol.\  {\bf 8}, 2 (2002)
  [hep-th/0104015].

\bibitem{Gibbons:2002tv} 
  G.~Gibbons, K.~Hashimoto and P.~Yi,
  ``Tachyon condensates, Carrollian contraction of Lorentz group, and fundamental strings,''
  JHEP {\bf 0209}, 061 (2002)
  [hep-th/0209034].

\bibitem{Kim:2011qh} 
  K.~-Y.~Kim, J.~P.~Shock and J.~Tarrio,
  ``The open string membrane paradigm with external electromagnetic fields,''
  JHEP {\bf 1106}, 017 (2011)
  [arXiv:1103.4581 [hep-th]].

\bibitem{Coleman:1976} 
  S.~Coleman,
  ``More about the Massive Schwinger Model,''
  Annals of Physics, {\bf 101}, 239-267 (1976).






\end{thebibliography}
\end{document}